# Square Net TaSiAs Nanowires with Topological Surface Conduction and Linear Magnetoresistance

Anand Roy*[a], Ofek Goldreich[a], Guy Ohad[a], Barun Barick[b] , Olga Brontvein[c], Ora Bitton[c], Katya Rechav[c], Yishay Feldman,[c] Leeor Kronik[a], and Ernesto Joselevich*[a]

[a]Department of Molecular Chemistry and Materials Science, Weizmann Institute of Science, Rehovot 76100, Israel. [b]Department of Condensed Matter Physics, Weizmann Institute of Science, Rehovot 76100, Israel. [c]Chemical Research Support, Weizmann Institute of Science, Rehovot 76100, Israel.

## Abstract

Square-net topological materials exhibit various interesting properties arising from the interplay of their structural diversity, topologically protected bands and different possible electronic features, including intrinsic magnetic ordering and superconductivity. However, little is known about the low-dimensional structures of these materials and their properties. Realization of low-dimensional topological square-net materials has the potential of enabling enhancement of their quantum behavior arising from quasi-compact 1D geometry and high surface-to-volume ratio, and their integration into functional devices. This work reports on the synthesis and properties of high-quality, single-crystal nanowires of Si square-net material TaSiAs. The chemical-vapor-transport growth produces TaSiAs nanowires that are chemically encapsulated *in situ* with a thin dielectric shell of $SiO_2$, enabling remarkable ambient stability and a pristine surface, which are critical for observing robust topologically protected surface states. Atomic-resolution structural analysis reveals a sharp core-shell interface, and a Si square-net lattice extending along the nanowire axis. Ohmic contacts fabricated on these chemically protected nanowires allowed us to observe rich electrical and magnetotransport features, including 4 to 11 times lower room-temperature resistivity than the closest bulk analogues and non-saturating linear magnetoresistance, revealing topologically protected coherent surface transport. First-principles calculations show a wide-range ($\geq$ 4 eV) linear band dispersion, alongside different Dirac cones that are protected by either symmorphic ($C_4$) or non-symmorphic symmetries, predicting transport features consistent with our results. The findings demonstrate the unique properties of low-dimensional square-net topological materials and their potential applications, including next-generation interconnects, spintronic devices and quantum computing.

## Introduction

A broad range of materials have been documented to contain two-dimensional (2D) square-net planes of atoms in their crystal structures.[1, 2] In these square-net materials, the alternating 2D square-net planes (4×4), stacked in the c-direction and separated by spacer layers, provide an interlayer coupling, giving rise to diverse structural and electronic properties.[1, 2] The charge transfer between the spacer layer and the square-net planes can influence structural stability. It has been shown that the overlap of square-net atomic orbitals (*s* and *p* orbitals) with the *d*-orbital of the spacer layer is crucial to the stability of 2D square-net planes, which otherwise undergo a Peierls distortion, transforming them into zigzag chains.[1] This structural change is often associated with the metallic/semimetallic phase transforming into semiconducting/insulating electronic states.[1]

Square-net materials with metallic/semimetallic electronic bands have received notable attention due to their potential to host topologically protected, wide-energy-range, linearly dispersive Dirac bands in the vicinity of the Fermi energy.[2] For instance, tetragonal (P4/nmm) MXZ compounds (M= Zr, Hf; X= Si, Ge, Sn; Z= S, Se, Te) wherein the 4×4 square-net (in the ab-plane) is constituted by Si/Ge/Sn atoms, offer nodal-line Dirac semimetallic bands with certain band crossings protected by the nonsymmorphic symmetry of the crystal.[2-4] The exotic band topologies in these materials give rise to exciting properties, such as low-effective-mass quasiparticles, high carrier mobility, substantially high and unsaturating magnetoresistance, quantum oscillations, and the half-integer quantum Hall effect.[2, 3, 5-9] Aliovalently substituted isostructural analogues of MXZ compounds, e.g., NbSiSb and NbGeSb with Si and Ge square-nets host an open Fermi surface with multiple electron and hole pockets that exhibits giant nonsaturating magnetoresistance.[10] Surface-sensitive photoemission spectroscopy revealed spin-polarized surface bands with Weyl-like cones and edge states in NbGeSb.[11] Structurally similar Dirac nodal-line LnSbTe compounds (Ln = lanthanides), with Sb-based square-net planes, can host strongly correlated electronic and antiferromagnetic states arising from the lanthanide *f*-orbitals of the spacer layer.[12-14] Orthorhombic $LnSb_2$ compounds (Ln = lanthanides) also host Sb-based 2D square-net planes and manifest complex magnetic ground states with anisotropic electrical transport.[15] $ABX_2$ (A = Ca, Sr, Ba, Eu, Yb; B = Mn; X = Bi, Sb) compounds can contain square-net planes composed of both Bi/Sb and Mn atoms, with antiferromagnetic ordering originating from the Mn-based 2D square-net planes.[2, 16] These compounds possess relatively complex band dispersions in the vicinity of the Fermi energy, with $SrMnBi_2$ displaying Dirac fermions with momentum-dependent Fermi velocities.[17, 18] The coexistence of magnetic ordering and exotic topologically protected bands can produce intriguing properties, such as the topological Hall effect and the anomalous Hall and Nernst effects.[16, 19, 20] For example, high-temperature ferromagnetic ordering with a record-high topological Hall effect was reported in the Mn-based square-net compound $LaMn_2Ge_2$.[21]

Most studies on the above-mentioned different families of square-net topological materials have been carried out using bulk single-crystals[2], while low-dimensional nanostructures of these materials are largely unknown. Low-dimensional topological materials, e.g., 1D nanowires (NWs), 2D flakes, and thin films can give rise to new or strongly enhanced quantum phenomena owing to the combined effects of quantum confinement and high surface-to-volume ratio.[22] A more ambitious goal could be generating Majorana zero modes (MZM) as end states at two ends in the intrinsically superconducting NWs or in the proximity of a s-wave superconductor.[23-26] These MZM in NWs have been proposed as the basis for topological quantum computers.[24, 26, 27] Recently, some of the present authors have demonstrated that topological semimetal $TaAs_2$ NWs chemically encapsulated with a $SiO_2$ shell show strongly enhanced quantum effects compared to bulk $TaAs_2$, including up to 15 times lower resistivity, giant magnetoresistance with metal-to-insulator transition occurring at four times higher temperatures than the bulk material, reaching room temperature, and a unique double pattern of Aharonov-Bohm oscillations (ABO).[28] All these observations are suggestive of coherent transport by topologically protected surface states. Besides the size reduction of the $TaAs_2$ in these NWs, the insulating $SiO_2$ shell around them effectively protects the $TaAs_2$ surface from chemical degradation at ambient conditions, thus allowing the observation of all these interesting quantum phenomena.[28] Extending this approach to square-net topological materials could analogously lead to a strong enhancement of their rich quantum properties, offering vast opportunities in nanoscience and nanotechnology.

Producing low-dimensional square-net materials could in principle be achieved by different methods. Given the ordered stacking of 2D square-net planes, a natural approach would be to produce nanoscale 2D structures either by mechanical exfoliation or by thin-film growth. Recently, a top-down mechanical exfoliation method was shown to produce thin flakes of $NdSb_2$; however, properties of the material at such low dimensions are still unknown.[29] Bottom-up synthesis approach, using chemical vapor precursors can offer high degree of freedom in controlling their structure with atomic-level control yielding minimum crystal defects.[30, 31] Motivated by our recent findings of $TaAs_2$ NWs[28] we attempted to synthesize NWs of TaAs, which is a prototypical type-I Weyl semimetal that exhibits giant bulk photovoltaic and giant nonlinear optical effects, owing to its bulk Weyl fermions and surface Fermi arc. A rational approach to achieve a TaAs phase was to use higher growth temperatures that may result in the thermodynamically favorable TaAs phase over $TaAs_2$. A higher growth temperature indeed produced high aspect ratio NWs, with the equiatomic Ta to As ratio expected for TaAs, and these NWs were chemically encapsulated by a thin $SiO_2$ shell similar to our $TaAs_2$-$SiO_2$ core-shell NWs.[28] To our surprise, however, we found that Si is not only part of the inert shell, but also present in the core alongside Ta and As, giving rise to a compound of composition TaSiAs. A careful examination of the structure of these NWs revealed them to have an ordered array of 2D Si-square lattices sandwiching the Ta-As spacer layers. Thus, the obtained compound TaSiAs shows a close resemblance to a compound one would obtain by sandwiching bilayer arrays of Weyl semimetallic TaAs,[32] between the square-net planes of Si.[33] Although both TaAs and TaSiAs crystallize in tetragonal structure, the presence of Si square-net planes adjacent to the Ta-As bilayer motifs renders the latter centrosymmetric (P4/nmm),[33] whereas the former remains noncentrosymmetric ($I4_1md$). On the other hand, the structural similarity between the TaSiAs family of compounds (NbSiAs, NbSiSb, NbGeSb, and NbGeAs)[10, 11] with the archetypical square-net containing MXZ compounds (where M= Zr, Hf; X= Si, Ge, Sn; Z= S, Se, Te) [2] is striking. Thus, TaSiAs nanowires offers an interesting platform to study the low-dimensional square-net family of materials that can manifest enhanced quantum and topological effects owing to a quasi-compact 1D geometry and a high surface-to-volume ratio.

In this work, we report on the NWs of the Si square-net compound TaSiAs, which, to the best of our knowledge, represent the first NWs, or any quasi-one-dimensional structure, documented for square-net topological materials. These NWs are synthesized by a high-temperature chemical vapor transport (CVT) route. Scanning (SEM) and transmission (TEM) electron microscopies, alongside energy dispersive X-ray (EDS) and Raman spectroscopy, have been used to examine their structure and composition. The TaSiAs NWs are *in situ* chemically encapsulated in a thin $SiO_2$ shell, yielding remarkably stable core-shell structures with a chemically protected TaSiAs surface. Atomically resolved TEM analysis of the NW core reveals that the Si square-net lattice extends along the NW axis. These NWs are integrated into four-terminal devices by local etching of the $SiO_2$ shell to make Ohmic contacts to the TaSiAs core, and their electrical and magnetotransport properties are studied. NWs were found to exhibit metallic conductivity with strong signs of topological surface-state-induced coherent electrical transport and Fermi-liquid behavior surviving up to a remarkably high temperature ($T \approx 100$ K). In thin NWs, a non-saturating linear magnetoresistance is observed in the presence of an external magnetic field applied perpendicular to the NW-axis. Density functional theory (DFT)-based first-principles calculations show different Dirac crossings protected by both symmorphic and non-symmorphic symmetry and linear band dispersions over a wide energy range ($\geq 4$ eV), which shed light on the measured transport properties.

## Results and Discussion

### Synthesis and Structure of TaSiAs Nanowires

TaSiAs crystallizes in the nonsymmorphic tetragonal space group P4/nmm.[33] The structure is composed of a Ta-As bilayer motif-based spacer layer, sandwiched between the 4×4 Si square-net planes stacked in the c-direction (**Figure 1a**). In the Ta-As bilayer motif, each Ta atom is coordinated with four As atoms within the layer and four Si atoms from the adjacent Si square-net layer, completing the coordination and thus forming a square antiprism coordination geometry around the Ta atoms (**Figure S1**). The Si square-net layers are composed of 4×4 square-net cells, wherein each Si atom is bonded with four nearby Si atoms in the same plane (**Figures 1a and S1**). The Ta-As spacer layer in TaSiAs resembles the structure of the prototypical type-I Weyl semimetal TaAs[34], with both sharing Ta atoms covalently bonded with four arsenic atoms within the layer; however, a structural difference arises due to the presence of Ta-Si covalent bonds in the former (**Figure S2**). This results in the latter being noncentrosymmetric ($I4_1md$)[34], whereas the former is centrosymmetric. On the other hand, the similarity between the TaSiAs structure and that of the heavily studied MXZ compounds[2], (M= Zr, Hf; X= Si, Ge, Sn; Z= S, Se, Te) places them in the same nonsymmorphic symmetry group (P4/nmm).[2] Note that reported square-net three-dimensional (3D) tetragonal crystals inherently prefer growth along their principal tetragonal axis i.e., the c-axis.[3, 35] However, realizing these materials in quasi-one-dimensional to one-dimensional forms may lower their symmetry and offer a competition between the preferential growth along the principal tetragonal axis and the square-net lattice plane. This anisotropic growth may give rise to one-dimensional structures of square-net materials. In **Figure 1b**, we show a graphical illustration of such anisotropic 1D TaSiAs NWs (Ta, As, and Si are represented by blue, yellow, and magenta atoms, respectively), which can be achieved by controlling crystal growth along one of the lattice vectors, parallel to the Si square-net lattice.

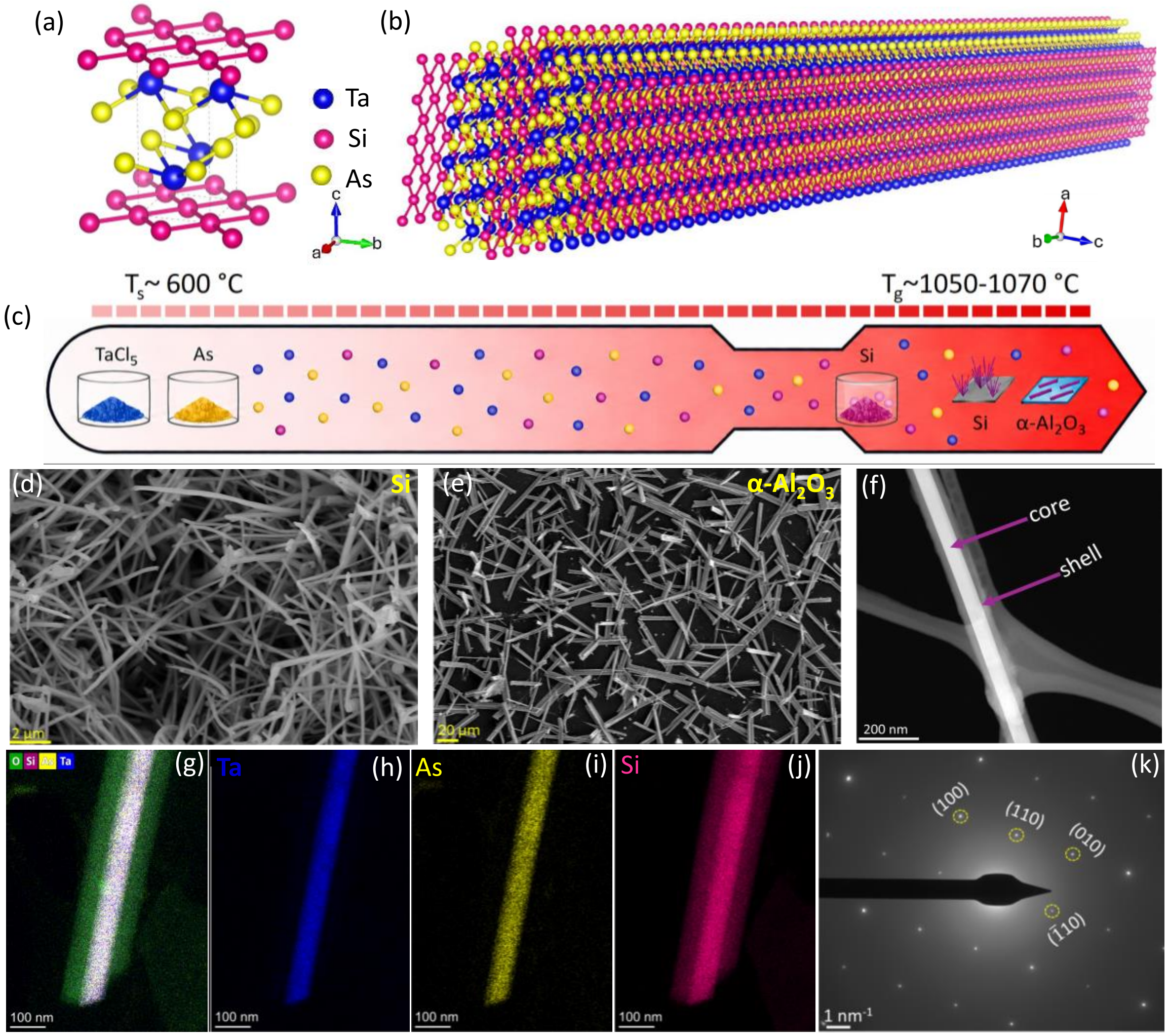

**Figure 1. Synthesis and structural analysis of TaSiAs/$SiO_2$ core-shell Nanowires (NWs)**. **(a)** Structure of the TaSiAs crystal (Ta, As, and Si atoms are shown in blue, yellow, and magenta). **(b)** Graphical illustration of a TaSiAs NW with the NW-axis along the Si square-net lattice. **(c)** Schematic illustration of the reaction tube and conditions involved. **(d)** SEM image of NWs grown on a Si substrate with a native oxide layer **(e)** SEM image of quasi-1D nano/micro-belts grown on an α-$Al_2O_3$ substrate. **(f)** HADDAF-STEM image of a NW showing TaSiAs-$SiO_2$ core-shell morphology. **(g-j)** STEM-EDS chemical maps of a NW displaying signals of Ta (blue), As (yellow), Si (magenta), and O (green). **(k)** Electron-diffraction (ED) obtained from the TaSiAs core with the [001] zone-axis.

We synthesized TaSiAs NWs inside a high-vacuum-sealed quartz tube, with a physical separation and thermal gradient between the precursor activation and the growth zone. A graphical illustration of the quartz ampoule containing the precursors and synthesis condition is shown in **Figure 1c**. The Si substrate with a native $SiO_2$ layer, placed at the junction of the precursor activation and growth zone (near the ampoule neck), acts as a source of Si. Growth in the temperature window of 1050-1070 °C, followed by quenching of the reaction to room temperature, yielded the Si substrates with dark grey films, which were revealed under a SEM as a high yield of NWs (**Figure 1d**). In addition to the Si substrate, adjacently placed $\alpha$-$Al_2O_3$ substrates separated by 1-2 mm also harvest NWs along with quasi-one-dimensional (Quasi-1D) nano/micro-belts (NBs) (**Figures 1e and S3**). Although a 5 nm thin layer of Au was deposited on the Si substrates for utilizing vapor-liquid-solid (VLS) growth of NWs[36], the absence of Au nanoparticles at the NW tips points to catalyst-free vapor-solid (VS) growth and not VLS.

A careful microscopic examination of NWs show them to have a uniform core-shell morphology, with the core of the NW displaying a diameter in the range of a few tens to a few hundred nanometers, and the shell around them maintaining a thickness of 50-60 nm (**Figure 1f**). A low-magnification high-angle annular dark-field (HAADF) STEM image of a core-shell NW, with its EDS chemical mapping, is shown in **Figures 1g-j**. The contrast between the core and shell, alongside the presence of Ta, Si, and As signals arising from the NW core, can be noticed. Note that in addition to the NW core the Si signal is also displayed from the NW shell. The STEM-EDS line profile shows modulations of atomic signals while moving from shell towards the core along the NW diameter, thus highlighting the core-shell nature of wires (**Figure S4**). Furthermore, the STEM-EDS spectra display NW core with a Ta:As atomic ratio close to 1:1 (spectrum 1 in **Figure S5**), and the shell with a Si:O ratio close to 1:2 (spectrum 2 in **Figure S5**). This indicates that the shell is made of $SiO_2$. Due to $SiO_2$ shell encapsulation of the NW measuring Si atomic fraction with respect to the Ta and As from the NW core becomes challenging (**Figure S5**). Nevertheless, the presence of equiatomic Ta to As ratio, alongside the distinct Si signals, strongly suggests a TaSiAs core. The electron diffraction (ED) pattern from the NW core, observed under the zone axis [001], is presented in **Figure 1k**; a square diffraction pattern originating from the identified planes, (100), (110), (010), and ($\bar{1}10$), confirms the single-crystalline nature of the NW core, and matches the TaSiAs structure.[33] Based on these microscopic observations, we can surmise that the NWs contain a single-crystalline TaSiAs core encapsulated in an amorphous $SiO_2$ shell.

To study the atomically resolved crystal structure of the TaSiAs core and its interface with the $SiO_2$ shell, electron-transparent lamellae of NW cross-section were produced (using a focused-ion beam (FIB)) and examined under TEM. A typical HAADF-STEM image of a NW cross-section manifests the irregular hexagonal shape of the crystalline TaSiAs core, which has parallel pairs of (001), (101), and ($10\bar{1}$) facets (**Figure 2a**), surrounded all around by an amorphous $SiO_2$ shell. NWs grown from a tetragonal crystal can be expected to have a preferred growth along the principal tetragonal axis, and preservation of the four-fold rotational symmetry is expected to produce a rectangular cross-section. The irregular hexagonal shape of TaSiAs crystal in our NW is noteworthy and indicative of growth along the [100] or [010] lattice directions (perpendicular to the c-axis). The STEM-EDS chemical mapping and quantitative spectra from the core and shell of the NW cross-section are shown in the inset of **Figure 2a** (bottom inset: tantalum (Ta) in blue, arsenic (As) in yellow, silicon (Si) in magenta, and oxygen (O) in green) and in **Figure S6**; these EDS images and analysis of their spectra clearly distinguish the composition of the core and the shell highlighting the sharp core-shell nature. The atomic-resolution HAADF-STEM image from the NW core reveals an ordered layered structure of TaSiAs (**Figure 2b**). In **Figure 2c** we present the atomically resolved HAADF-TEM image of the NW core overlaid with the atomic model of TaSiAs; resemblance between the structures is notable.

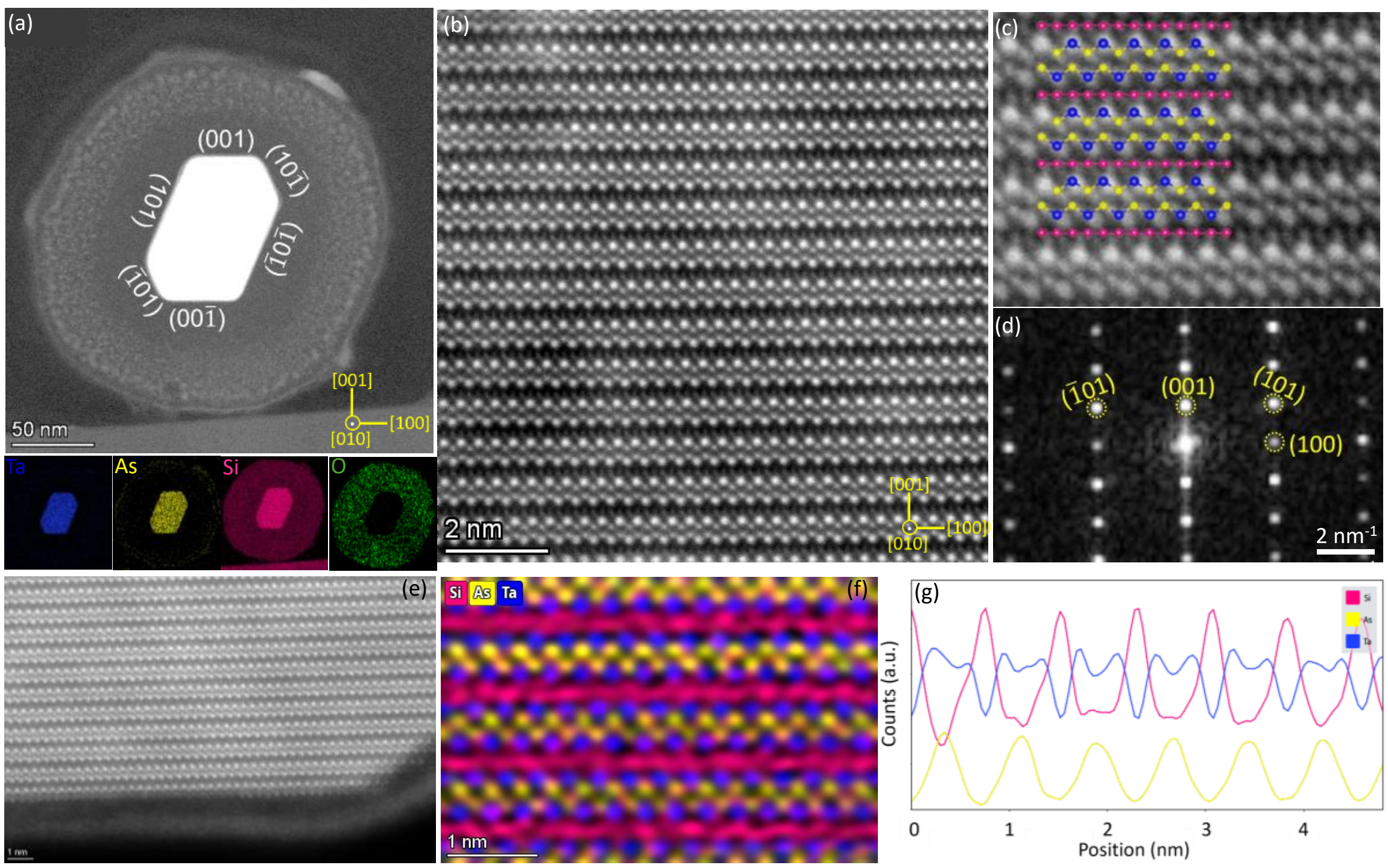

**Figure 2. Atomic-resolution crystal structure and EDS chemical mapping of core-shell TaSiAs-$SiO_2$ nanowire examined using an electron transparent thin NW cross-section. (a)** A low magnification HAADF-STEM image showing across-sectional core-shell view, exhibiting a crystalline and faceted TaSiAs core surrounded by an amorphous $SiO_2$ shell. Lower Insets in **a** show STEM-EDS chemical map of the core and the shell region (Ta, As, Si, and O signals are shown in blue, yellow, magenta, and green, respectively). **(b)** An atomic-resolution HAADF-STEM image obtained from the core displaying a TaSiAs structure. **(c)** Atomically resolved crystal structure of the core, along with TaSiAs atomic model overlaid. **(d)** Fast Fourier transformation (FFT) taken on the structure shown in **b,** presenting the reciprocal lattice of single-crystalline TaSiAs with lattice planes. **(e)** HAADF-STEM image showing a high-resolution core-shell structure from the interface region; a high quality TaSiAs surface can be noted. **(f)** Atomically resolved STEM-EDS chemical map from the TaSiAs core, showing Ta-As spacer layers (Ta (blue) and As (yellow) atoms) and Si square-net layers (magenta atoms). Ta-As bilayers sandwiched between Si square-net layers can be noted. **(g)** Corresponding EDS line intensity profile, showing periodic modulation of Ta, As and Si signals. Note that Ta and As signals are in phase and Si peaks are seen at a different lattice location.

The above structural analysis of the TaSiAs core confirms that the NW axis is along the [010] direction, which is parallel to the Si square-net lattice planes. The fast Fourier transform (FFT) of the image shown in **Figure 2b** displays a single-crystalline pattern consisting of (001), (101), (100), and ($\bar{1}$01) lattice planes (**Figure 2d**). An atomic-resolution HAADF STEM image showing the core-shell interface is presented in **Figure 2e**; a TaSiAs crystal interfacing with an amorphous $SiO_2$ shell while maintaining a high crystal quality and thus a pristine surface is observed. Atomically resolved EDS mapping of the NW core, showing Ta–As bilayer motifs (Ta in blue and As in yellow atoms) and Si square-net layers (magenta) between these bilayers units is presented in **Figure 2f**. The corresponding EDS line-profile mapping is shown in **Figure 2g**; the periodic modulations of the Ta (blue) and As (yellow) peaks and those of the Si peaks with different phase (magenta) at different lattice locations highlight the high crystalline quality of our NWs.

Note that quasi-1D nano/micro-belts (NBs) grown on $\alpha$-$Al_2O_3$ substrates also have a core-shell morphology, but unlike the single-crystal NWs, the TaSiAs core in these NBs is composed of different grains. In **Figures S7 and S8** we show a low-resolution cross-sectional TEM image, alongside EDS chemical mapping, from a ≈ 125 nm thick NB. A parallelogram shape of the NB cross-section with well-defined facets and a homogenous distribution of Ta, Si, and As can be noted. High-resolution TEM examination of an individual grain in this NB reveals that the bigger grain grows in the same <010> direction as the growth direction of NWs (**Figure S9**). The layers of the Si square-net and Ta-As motifs maintain the same stacking registries but are oriented at nearly a 50° angle from the NB axis (**Figures S9**). However, the smaller grain has a different growth direction i.e., the <$\bar{1}\bar{1}$1> direction of the TaSiAs crystal clearly manifested by a different FFT pattern of lattice fringes (**Figure S10**). The well-defined facets of the NBs forming a parallelogram shape are indicative of a single-crystal growth that later transforms into polycrystalline TaSiAs while maintaining the original single-crystal geometry. In a thicker NB, we observed that the TaSiAs core contains different density regions, suggesting different directional diffusion of Si atoms inside the original crystal (**Figure S11**). A plausible growth mechanism is that at the growth temperature shown in **Figure 1c,** the NB first grow as $TaAs_2$ single crystals and then undergo a chemical conversion to polycrystalline TaSiAs in the presence of Si vapor species, e.g., $SiCl_4$. Diffusion of Si atoms can occur through different facets of the $TaAs_2$ original crystal, thus producing different grains of TaSiAs while the original crystal's shape remains conserved. A conversion mechanism is further supported by the fact that the angles between the NB pseudofacets are very close to those of our previously reported $TaAs_2$ NWs (**Figure S11**).[28] Unlike these NBs, the TaSiAs NWs adopt the low-index facets (001), (101), and $(10\bar{1})$ (**Figure 2a**) of reported bulk TaSiAs, whose relative angles are different from those between the $TaAs_2$ NW facets. This difference between the facet/pseudofacets relative angles in the TaSiAs NWs and the NBs could be interpreted in two alternative ways: (i) The NWs grow directly as TaSiAs and not by the same conversion mechanism as the NBs. (ii) The NWs do form by the same conversion mechanism as the NBs but they faster reach the more stable facets of TaSiAs due to their smaller size. A thermodynamic preference for the TaSiAs phase over the binary phase $TaAs_2$, can be attributed to the enhanced entropy of mixing.[37]

In addition to the detailed microscopic structural analysis of the NWs and NBs, we carried out X-ray diffraction (XRD) on $1 \times 1$ cm$^2$ substrates covered with NWs and NBs. The resulting X-ray diffraction revealed all major peaks related to the TaSiAs phase (**Figure S12**). Single-NW-based Raman spectroscopy was used to determine the phase purity of NWs, as we found it beneficial for the quick examination of NW samples grown in different batches. Due to the absence of Raman spectral studies, we synthesized a polycrystalline phase of TaSiAs and characterized it using XRD (**Figure S13**). This high-purity phase TaSiAs polycrystal was used as a Raman spectral reference for our NWs and NBs. Although it was challenging to accumulate enough signal from very thin NWs, the relatively thicker NWs and NBs displayed Raman spectra with characteristic peaks at 92, 171, 208, and 420 cm$^{-1}$, which aligned with the reference spectrum of TaSiAs (**Figure S14**).

The use of a Si powder precursor instead of a Si substrate yields quasi-1D structures with a much lower aspect ratio than the TaSiAs NWs (**Figure S15a**). XRD examination of these samples revealed them to have a mixture of $TaSi_2$ (80%) and TaSiAs (20%) phases (**Figure S15b**). This emphasizes the importance of using a Si substrate with a native oxide layer as a source of Si to produce single-phase, high-quality TaSiAs NWs.

## Electronic Band Structure

In order to gain insight into the bulk and surface electronic structure, including possible topological protection, we performed first-principles calculations using the VASP[38, 39] software package (incl. the VASPKIT[40] utility for post-processing) to obtain the electronic band structure. The exchange-correlation energy was treated within the generalized-gradient approximation of Perdew-Burke-Ernzerhof.[41] The first Brillouin zone with high-symmetry lattice points of the primitive tetragonal TaSiAs crystal is shown in **Figure 3a.** In the absence of spin-orbit coupling (SOC) (**Figure 3b**), linear band crossings are observed along the **Γ-X**, **Γ-M**, **Z-R**, and **Z-A** paths. The crossings along the **Γ-X** and **Z-R** paths are located in very close vicinity of the Fermi energy ($E_f$), whereas the crossing along the **Γ-M** and **Z-A** paths lie slightly below $E_f$. In the absence of SOC, these band crossings are potentially protected by $C_{2v}$ symmetry, in close resemblance with the MXZ structural analogues (where M= Zr, Hf; X= Si, Ge, Sn; Z= S, Se, Te).[3, 4] Inclusion of SOC removes the linear crossings along these paths and opens gaps of 0.20 and 0.06 eV along the **Γ-X** and **Γ-M** direction, respectively (**Figure 3c**). These gaps are 10 and 3 times larger than the one reported in ZrSiS (20 meV)[3] and can be attributed to a stronger SOC from the Ta 5*d* orbital compared to Zr 4*d* orbital.

Note that with SOC included, the **Γ-M** path in TaSiAs maintains highly linear bands that crosses $E_f$ and maintains its linearity over a wide energy range from +2 to - 2 eV. The linearity and steep slope of this band imply a large group velocity, suggesting presence of high carrier mobility. The Dirac crossings at **X** and **R** points remain robust even after the inclusion of SOC and are located at 1.37 and 0.95 eV below $E_f$ respectively (**Figure 3c,** these crossings are highlighted in magenta dotted circles). This demonstrates the unique effect of the nonsymmorphic symmetry of square-net in TaSiAs that protects crossings at these points. The occurrence and locations of these nonsymmorphic symmetry-protected crossings resemble the ones observed in ZrSiS.[3] However, a striking difference between the TaSiAs and ZrSiS band topology appears along **Γ-Z** direction, where TaSiAs hosts a Dirac crossing that not only remains intact after including SOC, but interestingly moves nearly 100 meV closer to $E_f$, at ~ 0.18 eV below $E_f$ (**Figure 3c**, crossing is highlighted in green dotted circle). The robust feature of this Dirac cone in presence of SOC is indicative of its protection by the symmorphic $C_4$ symmetry of the TaSiAs crystal, deduced in analogy to a similar Dirac cone observation in tetragonal $Cd_3As_2$ ($I4_1cd$), a prototypical 3D Dirac semimetal.[42]

The orbital-projected band structures elucidate the atomic origin of band dispersions near $E_f$ (**Figure S16**). The wide-range linear band dispersion in the **Γ-M** path is dominated by the Ta 5*d* orbitals. The projected density of states (pDOS) near $E_f$ shows dominant contribution from the Ta 5*d* and As 4*p* orbitals with relatively smaller contribution from the square-net Si 3*p* orbitals (**Figure 3d**). Thus, the pDOS suggests a weak electronic coupling between the Ta-As spacer layer and the adjacent Si square-net planes, so we can expect electrical transport along the NW to be dominated by the Ta-As bilayers. On the other hand, the resemblance between the band topologies of TaSiAs and TaAs is limited, due to their fundamentally distinct symmetries. In TaAs, absence of inversion symmetry with strong Ta 5*d* based SOC leads to lifting of degeneracy and formation of chiral Weyl cones near $E_f$.[32] The presence of inversion and time-reversal symmetry in TaSiAs maintains Kramer degeneracy throughout the Brillouin zone and Weyl crossings are forbidden.

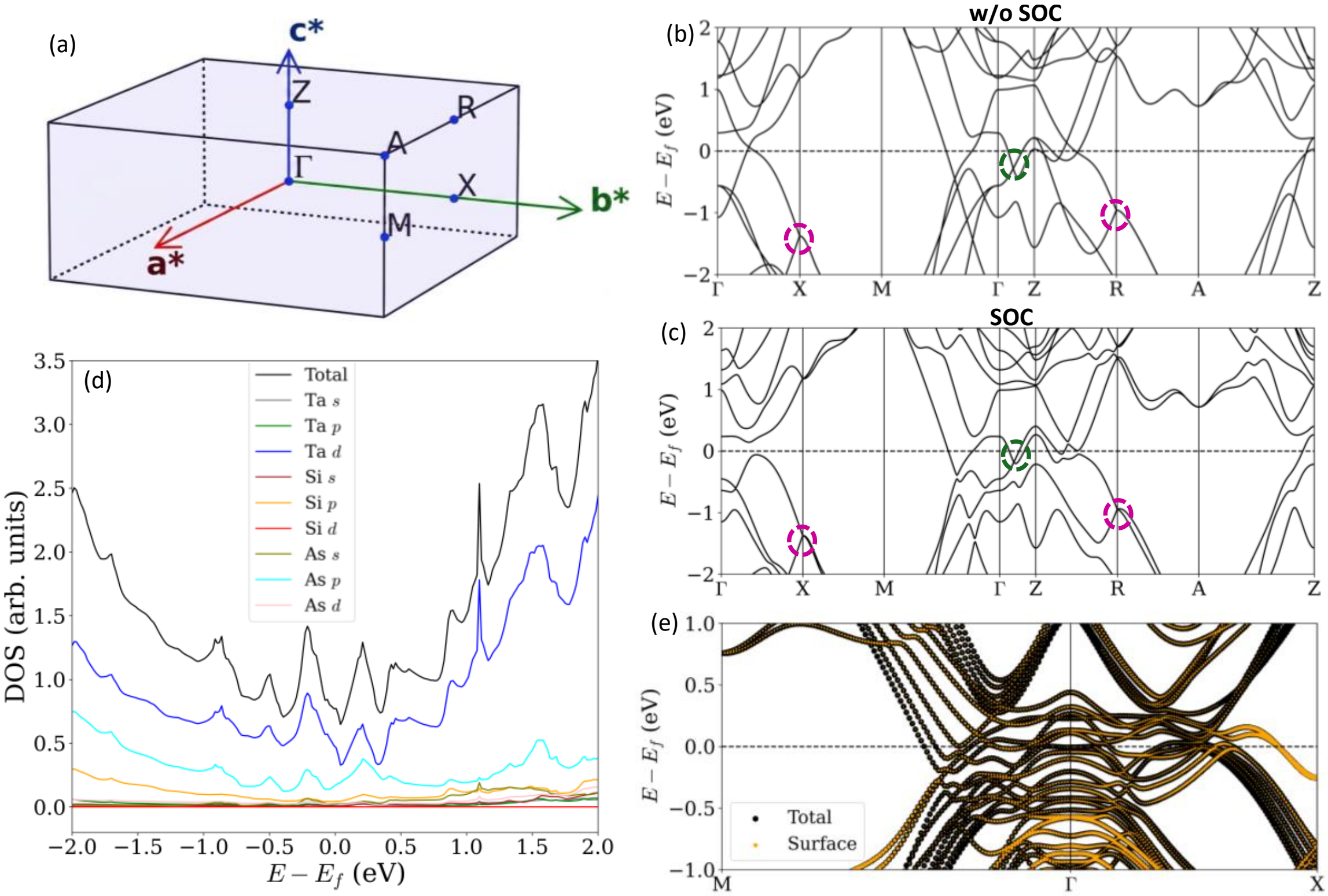

**Figure 3**. **Electronic band structure of bulk TaSiAs. (a)** First-Brillouin zone, with high-symmetry reciprocal lattice points indicated. **(b, c)** Electronic band structure without **(b)** and with **(c)** spin-orbit coupling, respectively. Symmetry-protected linear band crossings are shown by magenta (nonsymmorphic symmetry) and green ($C_4$ symmorphic symmetry) dotted circles. **(d)** Projected density of states (pDOS) with SOC included showing the elemental orbital contributions. **(e)** Calculated slab band structure, with SOC included, showing the (001) surface band dispersion. Yellow, and black dots represent surface and surface + bulk states respectively.

To visualize the surface electronic states, we performed a slab calculation by cleaving the two adjacent layers of As atoms to generate a (001) surface. **Figure 3e** presents the TaSiAs surface band structure of the (001) surface (the surface-projected spectral weight is indicated by yellow dots) including SOC. A dense distribution of surface-weighted bands spanning a broad energy range from approximately −1 eV to $E_f$ can be seen. Rather than a single dominant surface feature, multiple surface-derived branches are present, with substantial overlap between surface and bulk-projected states. It is also worth noting that the **Γ-X** path of surface Brillouin zone manifests gapless conducting states at $E_f$, while in the bulk there is a gap.

## Electrical transport Properties

The nanowires (NWs) and nano-/micro-belts (NBs) shown in **Figures 1e and 1f** were integrated into four-terminal devices (**Figures 4a and 4b**). Local etching of the $SiO_2$ shell through lithographically patterned openings followed by metal deposited created Ohmic contacts at specific locations on the TaSiAs core while maintaining chemical encapsulation over the rest of the wire (**Figures 4c and S17**, device fabrication details is provided in Methods section). In **Figure 4d**, we present the resistivity ($\rho$) as a function of temperature ($T$), measured for six different NWs with varying core diameters ($d$) of 75, 76, 80, 90, 125, and 620 nm. In all measured NW devices, the resistivity decreases upon cooling (**Figure 4d**). This temperature dependence of electrical transport is typical of metallic behavior, consistent with the electronic band structure shown in **Figure 3**, and aligns with the documented bulk analogue.[33] At 300 K, the resistivity ($\rho_{300\ K}$) of thin NWs ($d \approx 75$, 76, 80, 90, 125 nm) ranges from 100 to 130 μΩ cm, which decreases to $\rho_{2K}$ value between 10 to 54 μΩ cm at 2 K. In case of thicker NW ($d \approx 620$ nm) $\rho_{300\ K}$ and $\rho_{2K}$ values are 310 μΩ cm and 74 μΩ cm, respectively. Polycrystalline TaSiAs NBs also display metallic conductivity behavior with their resistivity decreasing with decreasing temperature (the NB device is shown in **Figure S18**). Resistivity as a function of temperature measured in two different NBs of thickness 255 nm (NB01) and 725 nm (NB02) is shown in **Figure 4e**. At 300 K, $\rho_{300K}$ for NB01 and NB02 is 154 and 398 μΩ cm respectively, which decreases to $\rho_{2K}$ of 25 and 78 μΩ cm at 2 K.

Several points are worth noting based on these electrical transport data that shed significant insight into the relative contribution of surface and bulk conduction. First, in both single-crystal NWs and polycrystalline NBs, the resistivity of thinner structures is notably lower than that of thicker analogues. For instance, the resistivity of thin NWs ($d \approx 75$, 76, 80, 90, 125 nm) is up to seven times lower than that of thicker NW ($d \approx 620$ nm) at 2 K, and three times lower at 300 K. Second, despite the polycrystalline nature of NBs (in **Figures S19 and S20** we show TEM images of the NB01 device cross-section, presenting different grains and orientations of Si square-net layer), the room temperature resistivity of the thin NB (NB01; $\rho_{300K}$ = 154 μΩ cm) is two times lower than that of the single-crystal thick NW ($d \approx 620$ nm; $\rho_{2K}$ = 311 μΩ cm ) and comparable to the resistivity values of single-crystal thin NWs ($\rho_{300K}$ =100 to 130 μΩ cm). Third, at low temperature (2 K), the resistivity of the thin polycrystalline NB (NB01; $\rho_{2K}$ = 25 μΩ cm) is up to three times lower than that of all measured single-crystal NWs ($\rho_{2K}$ = 49-74 μΩ cm), except for the NW with $d \approx 125$ nm that has a $\rho_{2K}$ value of 10 μΩ cm. In **Figures 4f** and **4g**, we show the resistivity scaling of NWs and NBs as a function of their cross-sectional area ($A$).

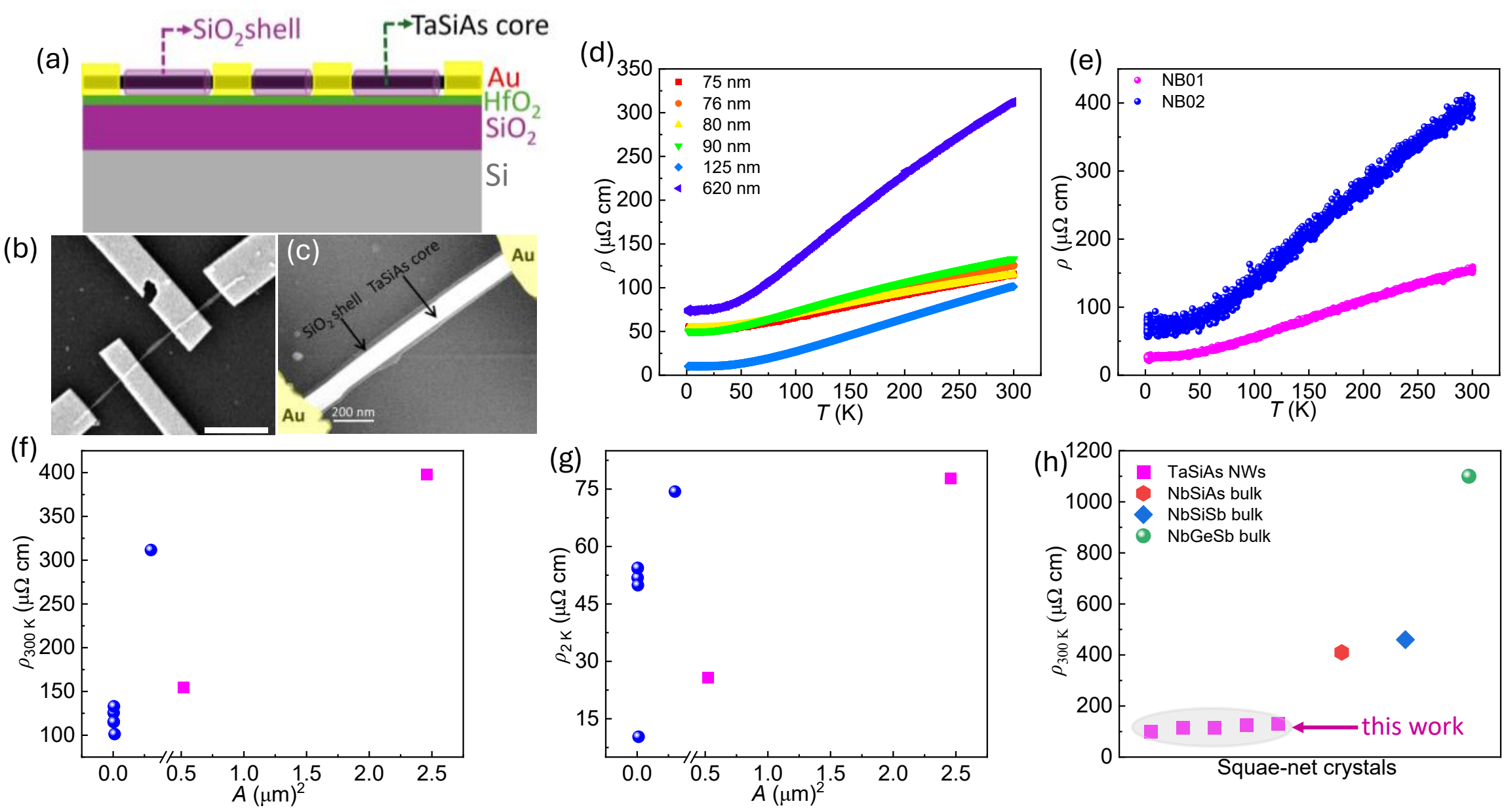

**Figure 4. Electrical transport properties of TaSiAs NWs. (a)** A schematic representation of four-terminal devices fabricated on a NW by local etching of dielectric $SiO_2$ shell, to make Ohmic contacts (Ti + Au shown in yellow squares) with TaSiAs core. (**b**) SEM image of a four-terminal NW-device. Scale bar 5 µm. (**c**) SEM image displaying local etching of $SiO_2$ shell to make Ohmic contacts with Au, while maintaining the shell encapsulation on TaSiAs core in between the electrodes. Scale bar 200 nm. **(d, e)** Electrical resistivity ($\rho$) as a function of temperature ($T$), measured in NWs of varying TaSiAs core diameter (**d**), and for quasi-1D NB of thickness 255 nm (NB 01) and 725 nm (NB 02) respectively (**e**). **(f, g)** Resistivity ($\rho$) as a function of cross-section area ($A$) for different nanostructures i.e., NWs in blue solid circles and NBs in purple squares at 300 K (**f**) and 2 K (**g**). (**h**) Comparison of room temperature resistivity ($\rho_{300K}$) values of TaSiAs NWs with the most closely related bulk square-net materials.

A general trend where a higher $A$ leads to higher resistivity in these structures can be noted at both 300 K and 2 K (**Figures 4f and 4g**). The fact that resistivity decreases with increasing surface-to-volume ratio suggests that conductance is dominated by surface states. Moreover, comparing the measured resistivity values of TaSiAs NWs with those of different bulk square-net analogues reported in the literature (given the fact that data for bulk TaSiAs are not available), further highlights the dominance of coherent surface transport in the NWs (**Figure 4h**). For instance, it is remarkable that the room-temperature resistivity ($\rho_{300K}$) of TaSiAs NWs is up to 4 times lower than that of its closest electronic and structural analogue NbSiAs ($\rho_{300K} \approx 410$ µΩ cm), which is a known superconductor with a critical temperature $T_c = 8$ K.[43] However, unlike NbSiAs, a transition to the superconducting state is not observed in TaSiAs NWs. Furthermore, the $\rho_{300K}$ of thin TaSiAs NWs is up to 5 and 11 times lower than those of NbSiSb ($\rho_{300K} \approx 460$ µΩ·cm) and NbGeSb ($\rho_{300K} \approx 1100$ µΩ cm) bulk single-crystals, respectively.[10]

More quantitatively, a temperature dependence of the resistivity as given by the empirical equation $\rho(T) = \rho_0 + AT^n$, is used to understand the scattering mechanism. Here, $\rho(T)$, $\rho_0$, $A$, $T$, and $n$ represent the measured resistivity as a function of temperature, residual resistivity, temperature-dependent scattering factor, temperature in Kelvin, and temperature exponent, respectively.[44] The temperature exponent ($n$) can have possible values between 1 and 5, arising from different scattering mechanism of charge carriers. Fitting of this equation for our single-crystal NWs and polycrystalline NBs in the low-temperature range ($T \approx$ 10–100 K), yields $n \approx 2$ (**Figure S21**). This suggests a Fermi-liquid behavior of electrical transport due to electron-electron scattering,[44, 45] which can be attributed to interband scattering caused by the small electron pockets near $E_f$ (**Figure 3c**).[45] In the high-temperature range ($T \approx$ 100–300 K), fitting yields $n \approx 1$, suggesting an electron-phonon scattering driven electrical transport (**Figure S21)**.[44, 45] This establishes that in both single-crystal NWs and polycrystalline NBs, electrical transport has a similar scattering mechanism. Thus, the superior electrical conductivity of thin NWs and NBs with respect to their thicker analogues at higher temperatures ($T \geq$ 100 K) can be attributed to the low electron-phonon scattering in the former owing to their high surface-to-volume ratio, which would enable a high surface-to-bulk transport ratio. At low temperatures i.e., 2 K, where electron-electron scattering dictates the transport, the 3 times enhanced conductivity in thin NB and up to 7 times enhanced conductivity in thin NWs compared with their thicker analogues suggests quasi-ballistic coherent surface transport which can arise from the electrical transport through topologically protected surface conducting channels, e.g., topological surface states.[28, 46]

The electrical conductivity of conventional metals decreases with decreasing their size due to enhanced surface and grain boundary scattering of carriers, an effect that become notably pronounced in nanoscale dimensions.[47, 48] However, in sharp contrast to conventional metals, in some topological semimetals it has been demonstrated that going from bulk to the nanoscale can increase conductivity, owing to the coherent surface transport from gapless Fermi arcs or Dirac electronic states.[28, 46, 49] The band structure of TaSiAs with topologically protected bulk Dirac cones, shown in **Figure 3c,** and the presence of gapless surface states (**Figure 3e**) are suggestive of such topological surface states in our NWs and NBs. In this case, the extreme surface-to-bulk ratio offered by quasi-1D thin NWs and NBs can give rise to strong manifestations of topological surface states. Thus, a thin NW or NB with high surface-to-volume ratio would show larger surface-to-bulk transport owing to the stronger manifestation of topological surface states. This topological surface state dominated transport also explains the superior electrical conductivity of a thin polycrystalline NB (NB01) with respect to the single-crystal NWs observed at low temperatures. In conventional polycrystalline metals the presence of grain boundaries acts as a scattering center for charge carriers, reducing their mean free path and thus giving rise to an enhanced resistivity.[50, 51] However, in topological materials electron backscattering at the surface is forbidden in the presence of time-reversal-symmetry due to spin-momentum locking of electrons;[26] thus, grain boundaries are less likely to affect the carrier mean free path. On the contrary, the exposed surfaces of different grains and grain boundaries may host exotic topological states which can further enhance surface conduction.[52] It is remarkable that our chemically encapsulated TaSiAs NWs and NBs manifest such strong signatures of topological surface conduction, which we attribute to the combined effects of a high surface-to-volume ratio and surface protection by the inert $SiO_2$ shell.

## Linear Magnetoresistance

Application of a magnetic field perpendicular to the TaSiAs NW axis induces magnetoresistance (MR), i.e., $(\rho-\rho_0)/\rho_0$; here, $\rho$ and $\rho_0$ are the resistivities in the presence and absence of an applied magnetic field ($B$), respectively. In **Figure 5a**, we present the field dependence of MR measured in a $d \approx 90$ nm NW device; note that the MR increases linearly with magnetic field and reaches ≈23% at 14 T, with no sign of saturation. In the inset (**Figure 5a**), we show the first derivative of MR, i.e., d(MR)/d$B$, as a function of field strength ($B$). The field-independent constant differential feature of the curve emphasizes the linear MR in this NW. Furthermore, observation of linear MR in two-terminal device configurations with longer channel length (**Figure S22a**), and its linear $B$ dependence on the measured temperature range of $T = 1.8$ to 50 K in both two- and four-terminal configurations, highlight its robustness (**Figures 5b and S22b** ). A careful analysis of the MR curve at low fields shows a crossover from a low-field quadratic ($m$ =2) dependence to the linear ($m$ =1) dependence at relatively larger fields (**Figure 5c**). The field at which the crossover takes place, i.e. the crossover field $B_c$, can be probed by a logarithmic plot shown in **Figure 5c**. $B_c$ does not show a clear temperature scaling but increases nearly three times in the measured temperature range of $T$ = 1.8 to 50 K and has values of 0.14 T and 0.34 T at 1.8 and 50 K, respectively (**Figure 5d**). At room temperature, the magnetoresistance still maintains the non-saturating feature but interestingly develops a super-linear field dependence, with a magnetic field exponent ($n$) of value 1.5 i.e., $MR \propto B^{1.5}$ (**Figure 5e**). This super-linear behavior at room temperature is likely to arise from a change in the scattering mechanism from low-temperature surface-dominated magnetotransport to high-temperature bulk-like phonon scattering.

To better understand this super-linear behavior, we studied magnetoresistance in NWs of different diameters ($d$) to probe the magnetic field exponent ($n$) as a function of NW diameter ($d$). It can be noted from **Figure 5f** that NWs with $d \leq 90$ nm manifest linear MR with exponent ($n$) = 1, whereas in thicker NWs ($d \geq 90$ nm) the exponent increases from $n = 1$ with its value being $n \approx 1.3$ and 1.6 for $d \approx 125$ and 620 nm NWs respectively. This diameter ($d$) dependent variation of magnetic field exponent ($n$) has been reported in NWs of trivial and topological materials, and was attributed to multiple factors such as the carrier cyclotron radius with respect to the NW radius, quantum interference from surface states, surface versus bulk magnetotransport, and characteristic magnetic field length with respect to the NW diameter.[28, 53-56] Bulk single-crystals of NbSiSb and NbGeSb exhibit parabolic nonsaturating MR.[10] Thus, we can suggest that in thin TaSiAs NWs at low temperatures, linear MR arises from surface-dominated magnetotransport, and deviation towards super-linear behavior in thicker NWs is due to the increasing contribution of magnetotransport from bulk cyclotrons. Magnetotransport from a different nanowire device ($d \approx$ 76 nm) is shown in **Figure S23**; similar linear MR features and their dependence on temperature can be noted.

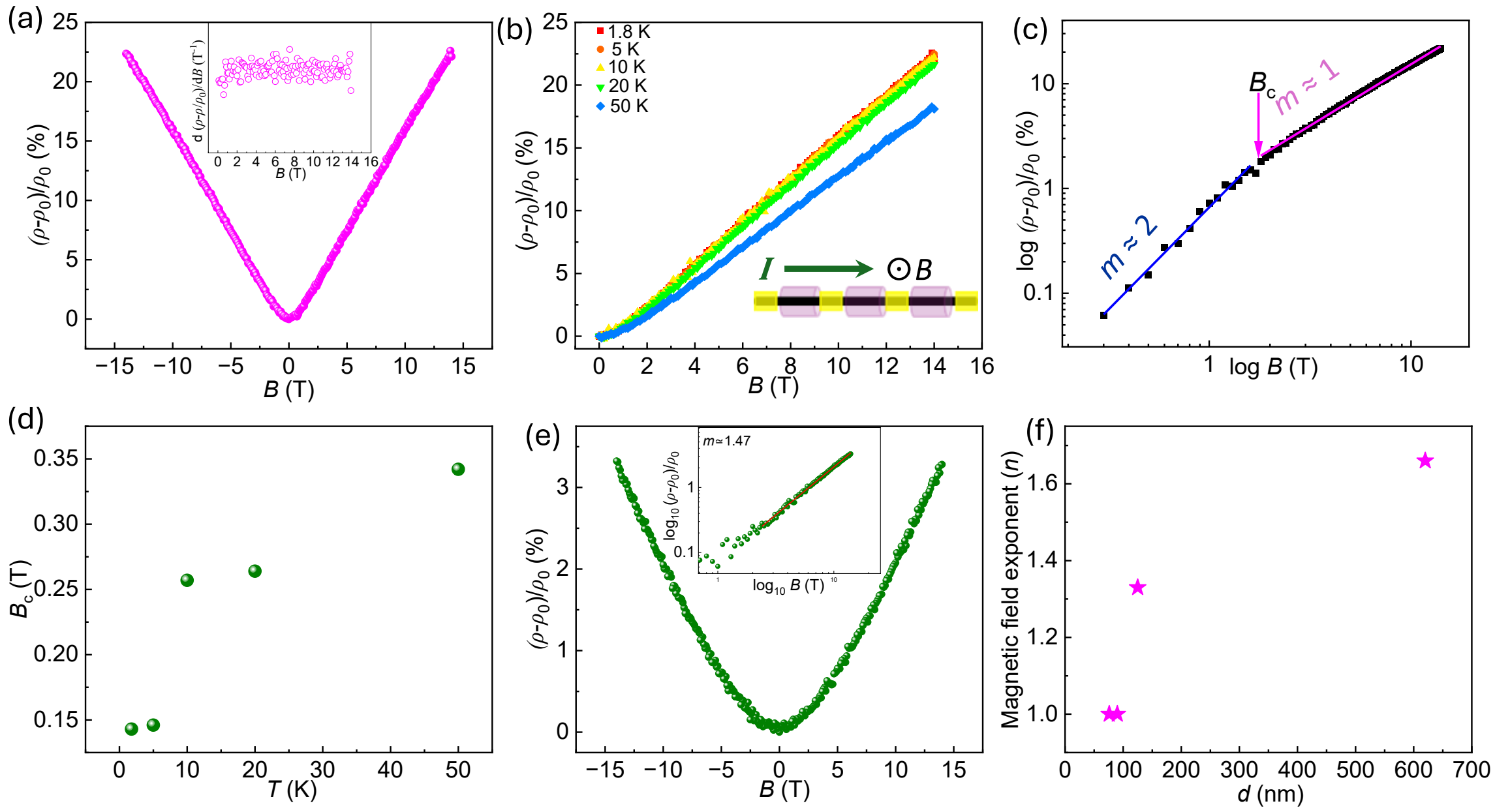

**Figure 5. Magnetotransport properties of TaSiAs NWs. (a)** Magnetoresistance (MR), $(\rho\text{-}\rho_0)/\rho_0$, as a function of magnetic-field strength ($B$), measured in a $d \approx 90$ nm NW at 2 K. Inset shows first derivative d(MR)/d$B$, with respect to the field strength ($B$). **(b)** MR as a function of magnetic field ($B$), at different temperatures for the same $d \approx 90$ NW device. Bottom right inset shows a schematic illustration of NW device (with locally etched $SiO_2$ shell) with direction of applied current ($I$) and magnetic field ($B$). **(c)** A $\log_{10}$ MR-$\log_{10} B$ plot and its fitting showing a switch of the MR feature from low-field quadratic ($m \approx 2$) to a linear ($m \approx 1$) MR at a crossover field ($B_c$) indicated by an arrow. **(d)** Temperature dependence of the crossover field ($B_c$) obtained in **c**. **(e)** MR as a function of field strength ($B$) at 300 K for the same $d \approx 90$ nm NW device. Inset shows $\log_{10}$ MR-$\log_{10} B$ plot with slope $m \approx 1.5$. **(f)** Magnetic field ($B$) exponent ($n$) as a function of TaSiAs core diameter ($d$) at 2 K.

The observation of linear MR and its robust nonsaturating feature extending over a broad field range (in sharp contrast to quadratic-to-saturating MR in conventional metals) has been observed in various materials, including narrow gap semiconductors[57], Dirac and Weyl semimetals[58, 59], topological insulators,[60] and quantum critical metals.[61, 62] It has maintained a sustained interest in both fundamental understanding of this phenomena in different systems and its potential technological advantages such as high-field magnetic sensing and magnetoelectric devices. In Dirac materials, the origin of linear MR has been attributed to their linear band dispersion.[63] TaSiAs hosts several linear bands near the Fermi energy, particularly a highly linear band along the Γ-M direction that crosses the Fermi energy ($E_f$) and maintains its linearity over a wide energy range of ≈ 4 eV (**Figure 3c**). Abrikosov's theory explains linear MR at the extreme quantum limit, where all carriers coalesce into the zeroth Landau level.[64] This requires a strong magnetic field to achieve $\hbar\omega_c > E_f$ at sufficiently low temperatures such that $\hbar\omega_c >> K_B T$ ; where $\hbar$, $\omega_c$, $E_f$, $K_B$ and $T$ are the reduced Planck constant, cyclotron frequency, Fermi energy, Boltzmann constant, and temperature, respectively.[64] The crossover field $B_c$ observed in our experiment has a much smaller value (**Figure 5d**) than the one expected to achieve Abrikosov's quantum linear MR. This makes Abrikosov's quantum linear MR explanation unsuitable in our case.

The classical origin of linear MR suggested by Parish and Littlewood based on random resistor networks requires a spatially varying current density due to an inhomogeneous mobility distribution arising from microscopic and macroscopic disorders.[65] In such systems, an average carrier mobility linked to the crossover field, $\bar{\mu}= 1/B_c$, is known to give rise the onset of linear MR, where $\bar{\mu}$ is the average mobility of carriers and $B_c$ is the crossover field defined above.[66] A high single-crystal quality of our NWs does not necessarily follow such a disordered sample picture. Nevertheless, it is possible to have nanoscopic defects and related electronic disorders that can give rise to classical linear MR. Considering the classical origin of linear MR in our NWs, yields an average carrier mobility ($\bar{\mu}$)= $1/B_c$ of 7.1 and 2.9 $m^2/V \cdot s$ at 1.8 and 50 K respectively, that leads to a carrier density of $1.7 \times 10^{24}$ $m^{-3}$ and $3.8 \times 10^{24}$ $m^{-3}$, at 1.8 and 50 K respectively (assuming a one-band model indicated by our band structure calculations and applying Drude's equation for the longitudinal conducvity $\sigma = ne\bar{\mu}$). This more than two-fold enhancement of carrier concentration in the temperature range of 2 to 50 K seems unlikely for a metallic system. Moreover, systems where linear MR has been attributed to a classical origin maintain the linear MR feature even at room temperature,[58, 67] unlike our observation where at room temperature we see a deviation from linear to super-linear MR behavior. Also, if the linear MR had a classical random resistor network origin, then polycrystalline TaSiAs NBs with higher inhomogeneity and disorder should manifest a stronger linear MR effect, whereas we observe a nonlinear MR (**Figure S24**). Thus, we can suggest that the linear MR in thin TaSiAs NWs is neither due to extreme Landau quantization (Abrikosov's model) nor due to inhomogeneous mobility (Parish-Littlewood model), but likely due to their magnetotransport being predominantly carried by surface states with linear energy dispersion, an effect that has been observed in low-dimensional topological Dirac materials with enhanced surface transport.[60, 68, 69]

## Conclusions

We reported the synthesis and properties of Si square-net TaSiAs nanowires (NWs), which, to the best of our knowledge, represent the first NWs, or any quasi-1D nanostructure, of square-net topological materials. The NWs are synthesized by chemical vapor transport and are *in situ* chemically encapsulated with a thin dielectric shell of $SiO_2$. This chemical encapsulation of the NWs, besides providing remarkable ambient stability and ease of device fabrication, effectively protects the surface from chemical degradation, enabling us to observe strong quantum effects arising from topological surface states. Quasi-1D nano/micro-belts (NBs) that grow under slightly modified synthetic conditions also feature a core-shell structure wherein a thin dielectric $SiO_2$ shell encapsulates the core, but the TaSiAs core here is polycrystalline. Atomically resolved structural examination of the TaSiAs core revealed the same stacking registry of the Si square-net and TaAs layers in both NWs and NBs, with the latter displaying a tilted orientation of the Si square-net layers with respect to the [001] axis. A plausible growth pathway is a conversion of single-crystalline binary compound (e.g. $TaAs_2$) NW into TaSiAs NWs and NBs, favored by a high entropy of mixing. Four-terminal devices produced by locally etching the dielectric $SiO_2$ shell, while preserving chemical encapsulation around the rest of the NW/NB core, enabled the observation of electrical and magnetotransport features strongly expressed by their pristine surface states. Electrical transport shows a Fermi-liquid behavior persisting up to remarkably high temperatures ($T \approx 100$ K), and a strong manifestation of coherent transport by topological surface states. The grain boundaries in thinner polycrystalline NBs are not detrimental to electrical transport; on the contrary, they are likely to enrich topological surface states, making the electrical transport comparable and in some cases even superior to that of single-crystal NWs. These findings highlight the robustness of topologically protected surface conduction channels and suggest their potential use as next-generation circuit interconnects and thermoelectrics for nanoscale devices. In a magnetic field perpendicular to the NW axis, a nonsaturating linear magnetoresistance is observed, which is attributed to the transport by gapless surface states. First-principles calculations show that TaSiAs hosts multiple Dirac cones, some of which remain ungapped even in the presence of spin-orbit coupling. The Dirac cones at the Brillouin zone boundaries (**X** and **R** points) are protected by the nonsymmorphic symmetry of the crystal, resembling those of ZrSiS and its analogues. A Dirac cone on the **Γ**-**Z** path lies 0.18 eV below the Fermi energy and is protected by the symmorphic $C_4$ symmetry, which can be easily accessed by a slight *p*-doping. A highly linear band extending over a wide energy range of ≥ 4 eV is also noted. In a broader vision, our findings open the possibility of synthesizing NWs and quasi-1D nanostructures of other square-net topological materials, including the close superconducting analogue NbSiAs, which could be an interesting topological superconductor candidate for observing Majorana Fermions and related phenomena. These novel 1D to quasi-1D structures of square-net quantum materials, alongside their *in situ* chemical protection, can show enhanced quantum properties and hence are promising building blocks for various developing technologies, including next-generation interconnects, spintronics, thermoelectrics and quantum computing.

## Methods

### Synthesis

TaSiAs NWs and NBs were synthesized in a custom-made quartz ampoule with physical separation between the precursor activation and nanowire growth zones (**Figure 1c**). $TaCl_5$ (s) and As (s) powder in an equimolar ratio (or with an excess of As) were placed at the bottom end of the quartz ampoule, and Si substrates with native $SiO_2$ (1-2 nm) layers, along with different planes of α-$Al_2O_3$ substates were placed in the growth end of the ampoule with a 20-21 cm separation from the powder precursors (**Figure 1c**). This ampoule was evacuated under a vacuum ($2\times10^{-5}$ mbar) and then sealed. The sealed ampoule was placed in a two-zone heating horizontal tube furnace. During the synthesis, the precursor end of the ampoule was maintained at nearly 550 °C (± 15 °C), and the growth end of the ampoule was maintained between 1050-1070 °C for 20 minutes, followed by quenching the reaction from the growth temperature to room temperature. The Si substrate with a native oxide layer placed in the growth zone serves as the source of Si, potentially by converting into $SiCl_4$ in the presence of liberated $Cl_2$ at high temperatures. Among different halide precursors, $TaCl_5$ results in the best yield of NWs and NBs. A plausible growth mechanism for the $SiO_2$ shell around the TaSiAs NW and NB core is likely to be the one suggested by us for $TaAs_2$ NW.[28]

### Characterization

The morphology of NWs and NBs was examined in a scanning electron microscope (Sigma 500 SEM, Zeiss). The core-shell structure of was examined in a high-magnification SEM (Sigma 500 Zeiss) at an accelerating voltage of 10 KV and above. It was further confirmed using TEM (Talos F200X G2 TEM). Thin electron-transparent lamellae were produced using focused-ion-beam (FIB, FEI Helios 600, Dual-beam microscope). The atomic resolution crystal structure analysis of nanowire was done using a high-resolution transmission microscope (aberration corrected S/TEM, Themis-Z). Raman spectrum of the samples was recorded using a micro-Raman spectrometer (Horiba LabRAM HR evolution). A 532 nm green laser was aligned and focused on a single NW/NB specimen under an optical resolution of X100.

### Device Fabrication and Transport Measurements

Si/$SiO_2$ (285 nm) substrates were capped with a 10 nm layer of $HfO_2$, deposited by atomic layer deposition (ALD). These substrates were treated with acetone + isopropanol + water and dried at 150 °C. Nanowires were transferred to the treated Si/$SiO_2$/$HfO_2$ substrate using a PDMS stamp and spin-coated with a photoresist (S1813) or electron beam resist (PMMA A5). Using a laser or electron beam writer, 1-2 µm size rectangular windows were created on specific locations of core-shell nanowires. This followed the wet etching (BOE 6:1; 6:1 volume ratio of 40 % $NH_4F$ in $H_2O$ to 49 % HF in $H_2O$) of $SiO_2$ shell (20-30 seconds) through lithographically created windows to expose the TaSiAs NW core. The core was contacted with 5 nm Ti and 250 nm Au electrodes using an electron beam evaporator at room temperature. The use of ALD-$HfO_2$ capping layer on Si/$SiO_2$ protects the $SiO_2$ layer of the substrate against BOE etching. The magnetotransport measurements were performed in a Quantum Design 14 T PPMS Dynacool system, using the electrical transport and the resistivity modes. These options generate either a DC or an oscillating low-frequency current and measure the voltage across the device using a four-probe wiring scheme. Currents of up to 100 nA were used, with frequencies around 0.5 Hz in the electrical transport option.

## Associated Content

### Supporting Information

Supporting Information is available

Structural comparison of TaSiAs and TaAs; SEM and TEM images alongside the EDS chemical mapping of TaSiAs-SiO2 core-shell nanowires (NWs) and nano/micro-belts (NBs); Powder X-ray diffraction of TaSiAs NWs and NBs; Power X-ray diffraction pattern of bulk TaSiAs powder, along with the Raman spectra of NBs and bulk; SEM image and powder X-ray diffraction of reaction product obtained using Si-powder precursor; The orbital projected band structure of TaSiAs; SEM and AFM images of NW and NB devices; Cross-sectional TEM of NB01 device; Fitting of resistivity as a function of temperature curve; Magnetoresistance plots for two-terminal devices; Magnetotransport properties of a 76 nm diameter NW device; Magnetoresistance plot for a NB01 device.

## Author Information

### Corresponding Authors

Ernesto Joselevich -Department of Materials Science and Molecular Chemistry, Weizmann Institute of Science, Rehovot 76100 Israel; E-mail: ernesto.joselevich@weizmann.ac.il

Anand Roy- Department of Materials Science and Molecular Chemistry, Weizmann Institute of Science, Rehovot 76100 Israel; Email: anand-kumar.roy@weizmann.ac.il

## Acknowledgments

A.R. acknowledges Faculty Dean's postdoctoral fellowship. We thank Dr. Iddo Pinkas for assistance with Raman spectroscopy, Dr. Eran Mishuk for help with ALD, and Prof. Reshef Tenne for providing synthetic facilities and knowledge. We thank Marcus Hucker for providing transport measurement facilities. A.R. thanks Dr. Mukesh Kumar Dasoundhi and Dr. Shashank Chaturvedy for several fruitful discussions. A.R. thanks Ms. Ewa Polinska for technical support in writing this manuscript and Bibliotika Gdynia (Poland) for providing wonderful atmosphere to write this manuscript. This research was supported by the Israel Science Foundation (ISF) grant No. 1045/23. E.J. holds the Drake Family Professorial Chair of Nanotechnology. L.K. thanks the Mintzi and Aryeh Katzman Professorial Chair and the Helen and Martin Kimmel Award for Innovative Investigation for their support. Computational work was performed at the CHEMFARM HPC facility of the Weizmann Institute of Science, supported in part by the Ben May Center for Chemical Theory and Computation.

## Conflict of Interest

The authors declare no conflict of interest

## Data Availability Statement

The data that support the findings of this study are available from the corresponding author upon reasonable request.

**TOC**

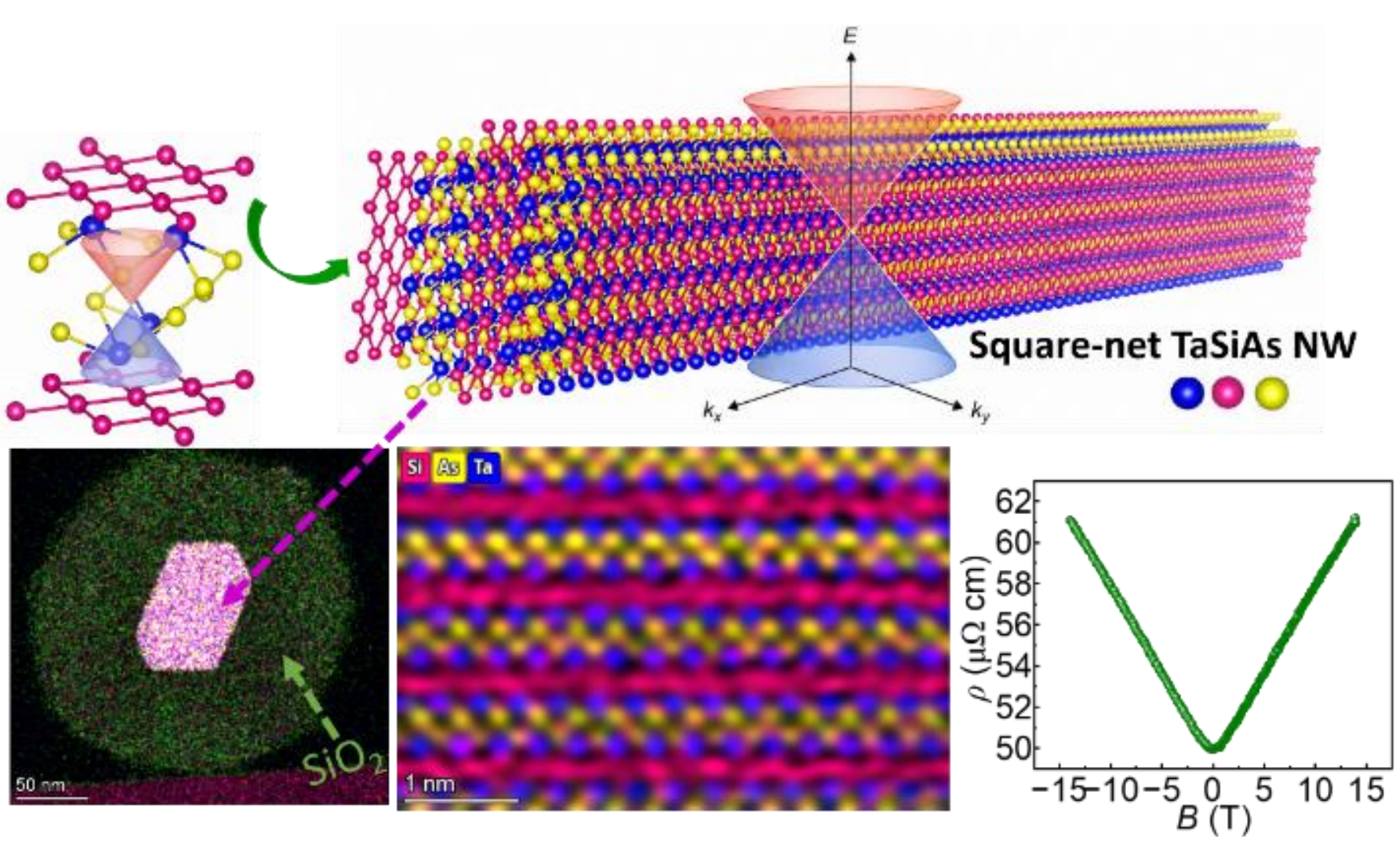

Supplementary Information

for

# Square Net TaSiAs Nanowires with Topological Surface Conduction and Linear Magnetoresistance

Anand Roy*[a], Ofek Goldreich[a], Guy Ohad[a], Barun Barick[b], Olga Brontvein[c], Ora Bitton[c], Katya Rechav[c], Yishay Feldman[c], Leeor Kronik[a], and Ernesto Joselevich*[a]

[a]Department of Molecular Chemistry and Materials Science, Weizmann Institute of Science, Rehovot 76100, Israel. [b]Department of Condensed Matter Physics, Weizmann Institute of Science, Rehovot 76100, Israel. [c]Chemical Research Support, Weizmann Institute of Science, Rehovot 76100, Israel.

**Content**

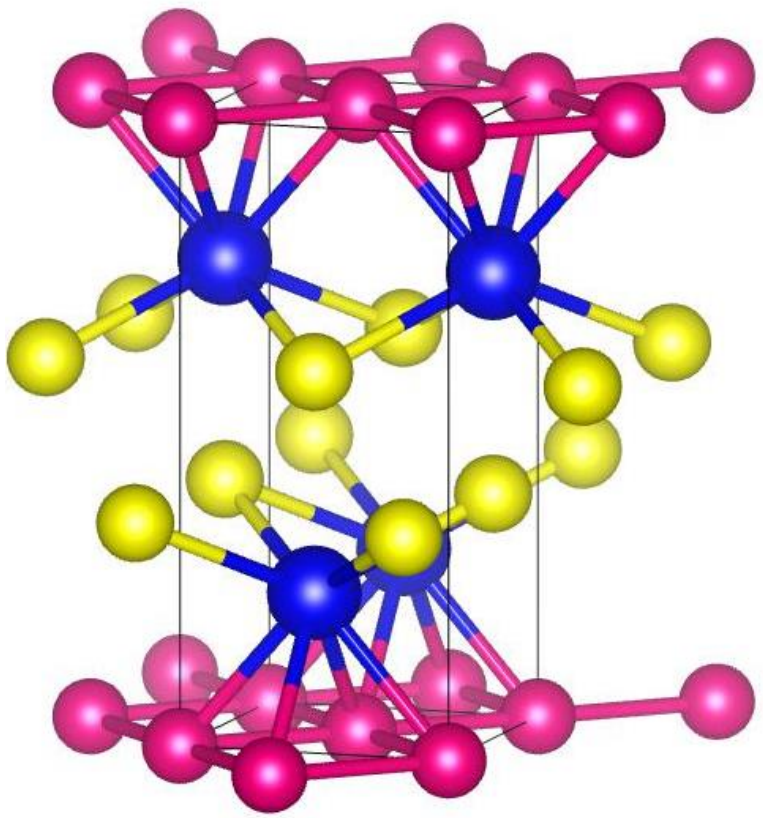

**Figure S1.** Unit cell of TaSiAs showing a Ta-As bilayer sandwiched between the square net lattice of Si. Si, Ta, and As are represented by magenta, blue and yellow atoms. Note that Ta atoms in a Ta-As motif are coordinated with four arsenic atoms and four Si atoms from the adjacent Si square net layer resulting in a square antiprism coordination geometry of Ta.

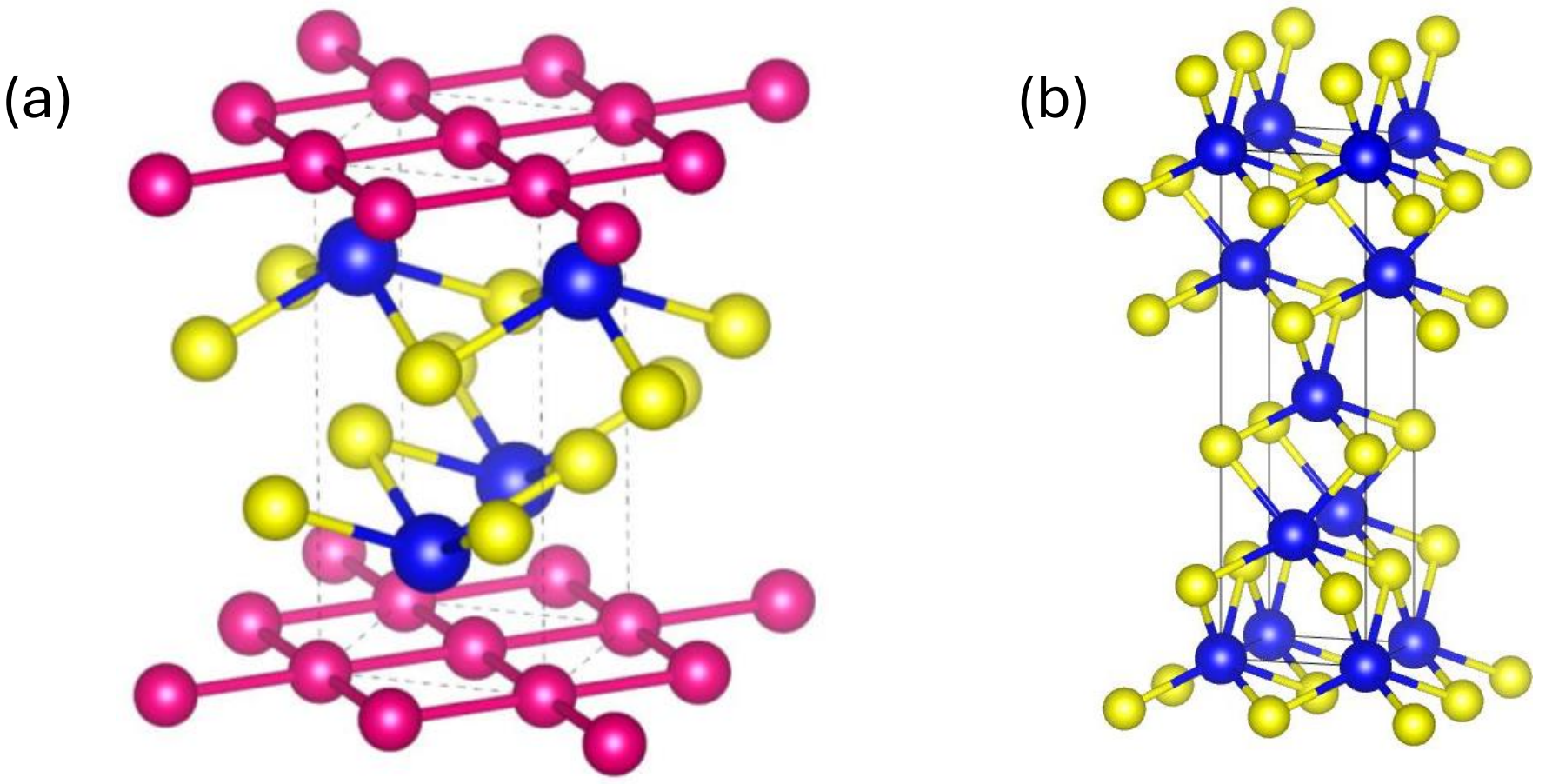

**Figure S2.** Structural similarity between (**a**) TaSiAs and (**b**) the prototypical Weyl semimetal, TaAs

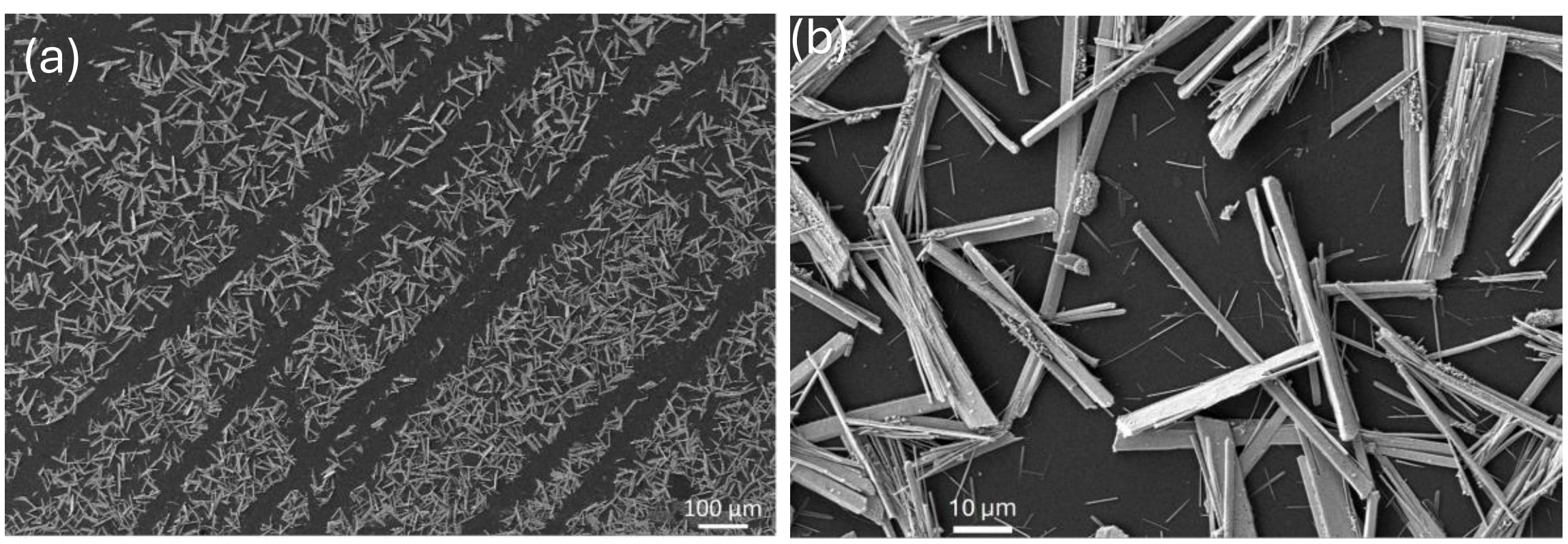

**Figure S3.** (a) Low-magnification SEM image showing large area growth of TaSiAs NBs. (b) TaSiAs NBs shown under relatively high-magnification SEM.

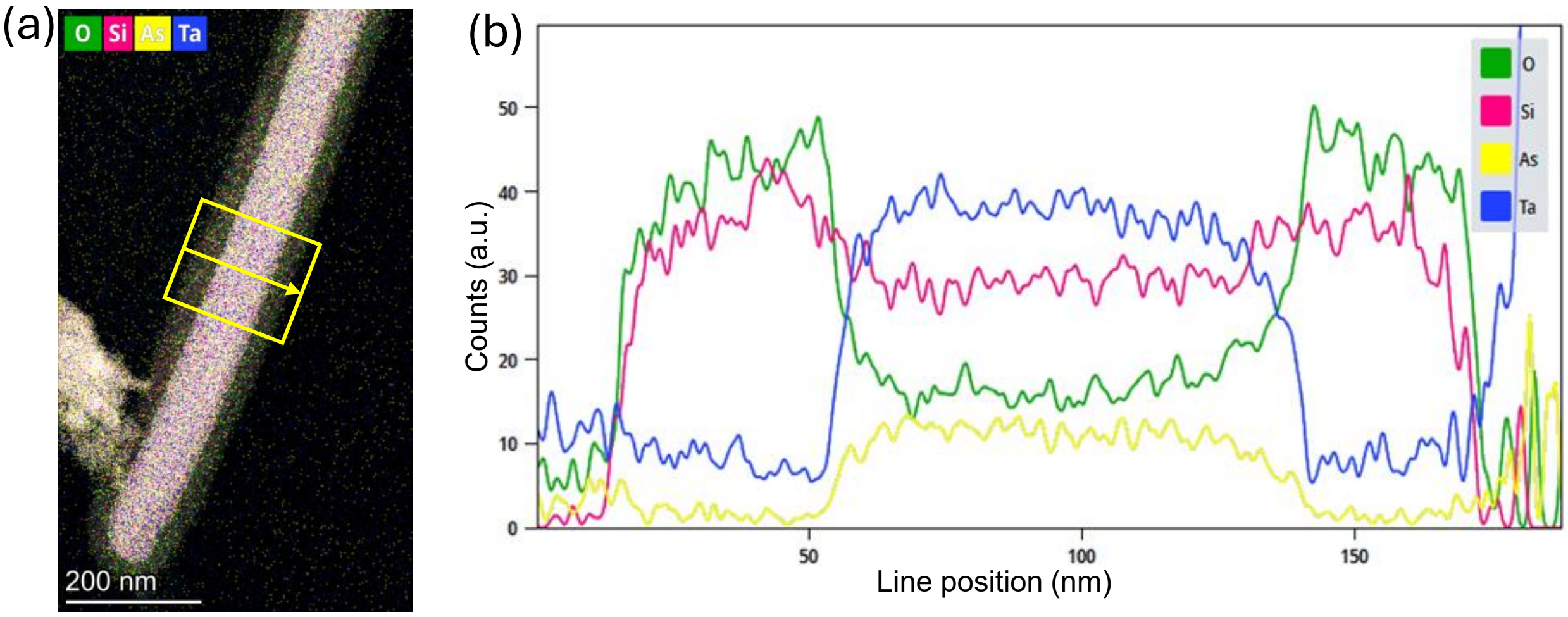

**Figure S4.** A typical EDS line-profile highlighting the core-shell morphology of TaSiAs-$SiO_2$ NWs.

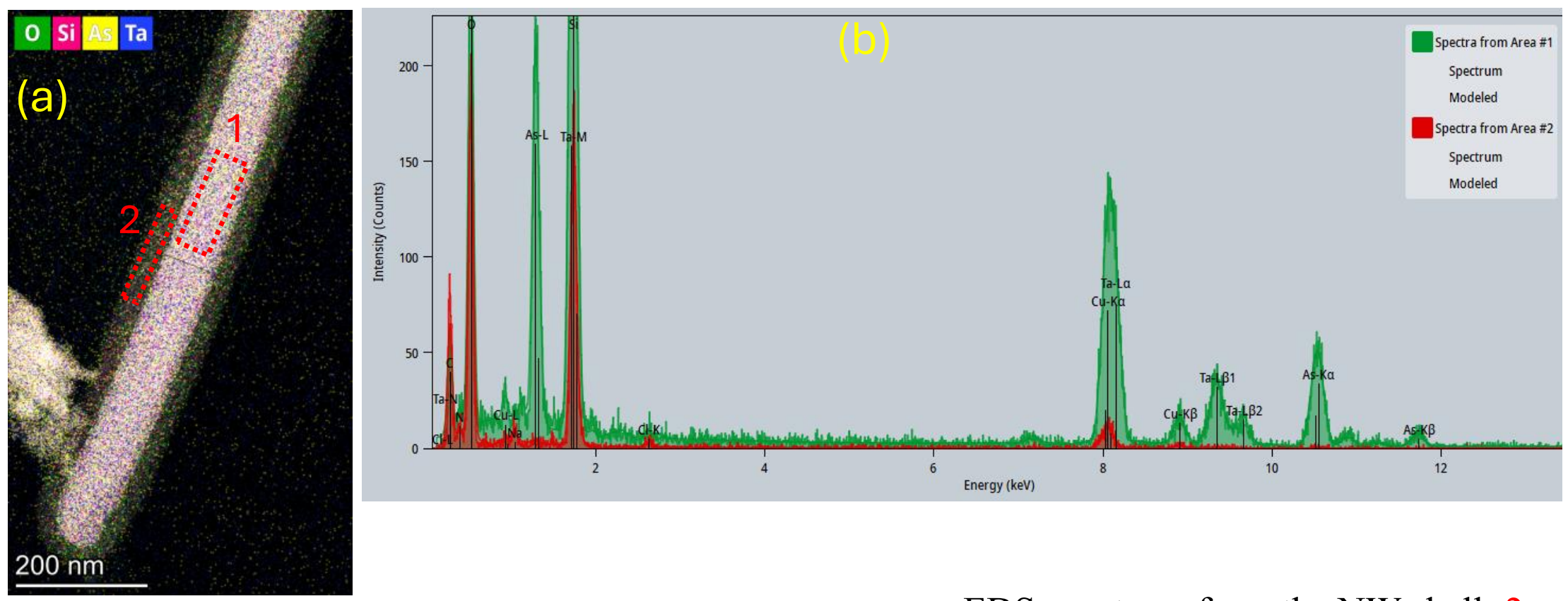

(c) EDS spectrum from the NW core: 1

| Z | Element | Family | Atomic Fraction (%) | Atomic Error (%) |
|---|---|---|---|---|
| 7 | N | K | 2.51 | 0.30 |
| 8 | O | K | 44.44 | 5.82 |
| 11 | Na | K | 0.44 | 0.12 |
| 14 | Si | K | 33.46 | 5.34 |
| 17 | Cl | K | 0.53 | 0.12 |
| 33 | **As** | **K** | **9.53** | 1.56 |
| 73 | **Ta** | **L** | **9.09** | 1.35 |

(d) EDS spectrum from the NW shell: 2

| Z | Element | Family | Atomic Fraction (%) | Atomi Error (' |
|---|---|---|---|---|
| 7 | N | K | 2.82 | 0.44 |
| 8 | **O** | **K** | **61.58** | 6.23 |
| 11 | Na | K | 1.48 | 0.37 |
| 14 | **Si** | **K** | **32.87** | 5.97 |
| 17 | Cl | K | 0.65 | 0.16 |
| 33 | As | K | 0.23 | 0.07 |
| 73 | Ta | L | 0.38 | 0.07 |

**Figure S5. STEM-EDS analysis of a NW**. (a) Image showing EDS mapping of a NW with a region from the core (1) and shell (2) highlighted in red dotted square. (b) EDS spectra comparing the core (1) and shell (2) region in image (a). (c) Corresponding atomic percentage from the core (1) and shell (2) respectively.

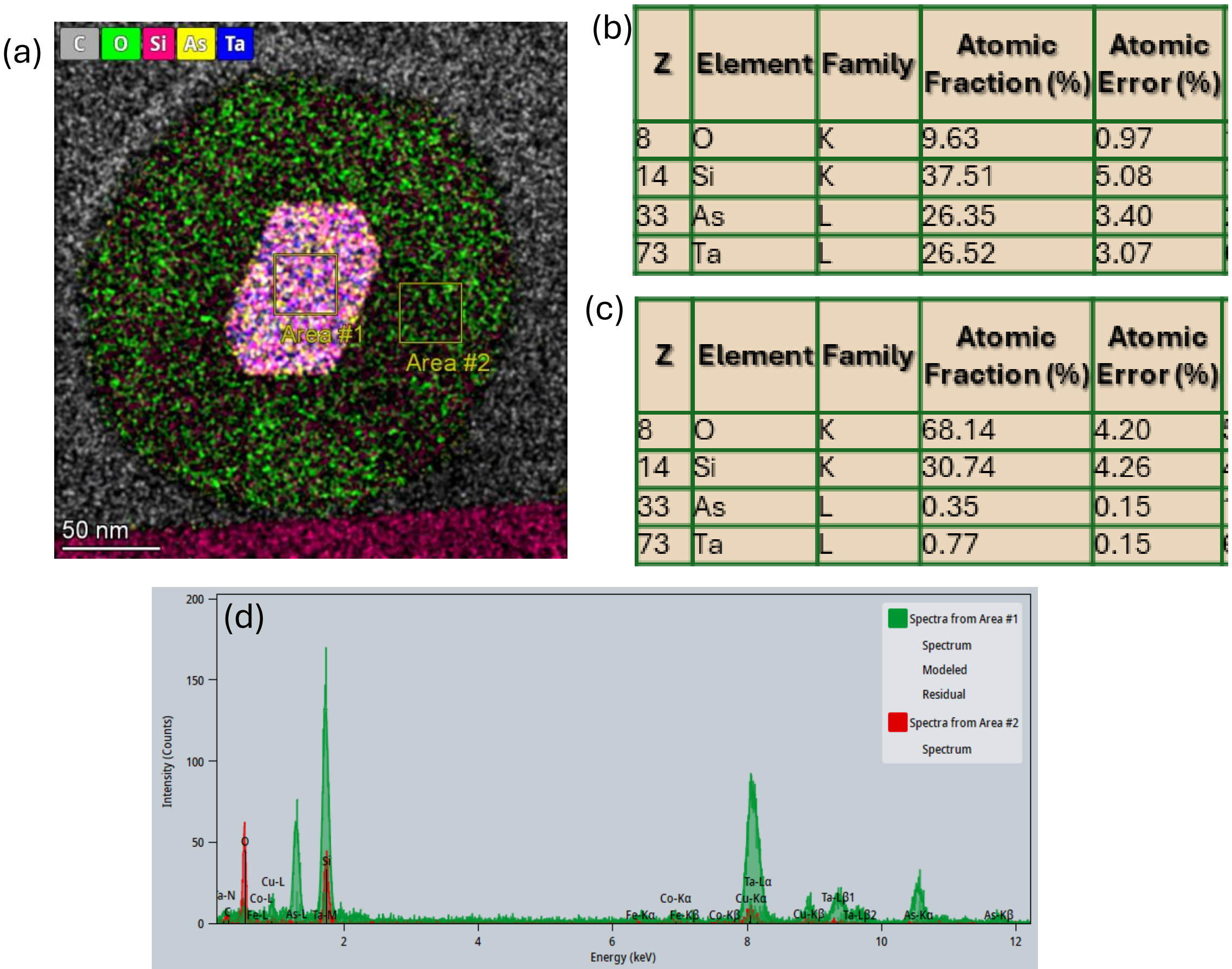

(b)

| Z | Element | Family | Atomic Fraction (%) | Atomic Error (%) |
|---|---|---|---|---|
| 8 | O | K | 9.63 | 0.97 |
| 14 | Si | K | 37.51 | 5.08 |
| 33 | As | L | 26.35 | 3.40 |
| 73 | Ta | L | 26.52 | 3.07 |

(c)

| Z | Element | Family | Atomic Fraction (%) | Atomic Error (%) |
|---|---|---|---|---|
| 8 | O | K | 68.14 | 4.20 |
| 14 | Si | K | 30.74 | 4.26 |
| 33 | As | L | 0.35 | 0.15 |
| 73 | Ta | L | 0.77 | 0.15 |

**Figure S6.** (a) STEM-EDS mapping from the core and shell of a NW lamella. The obtained atomic ratio from **(b)** the core and (**c**) the shell of a NW cross-section. (d) Corresponding EDS spectra from the core and shell regions highlighted in yellow square boxes in (a).

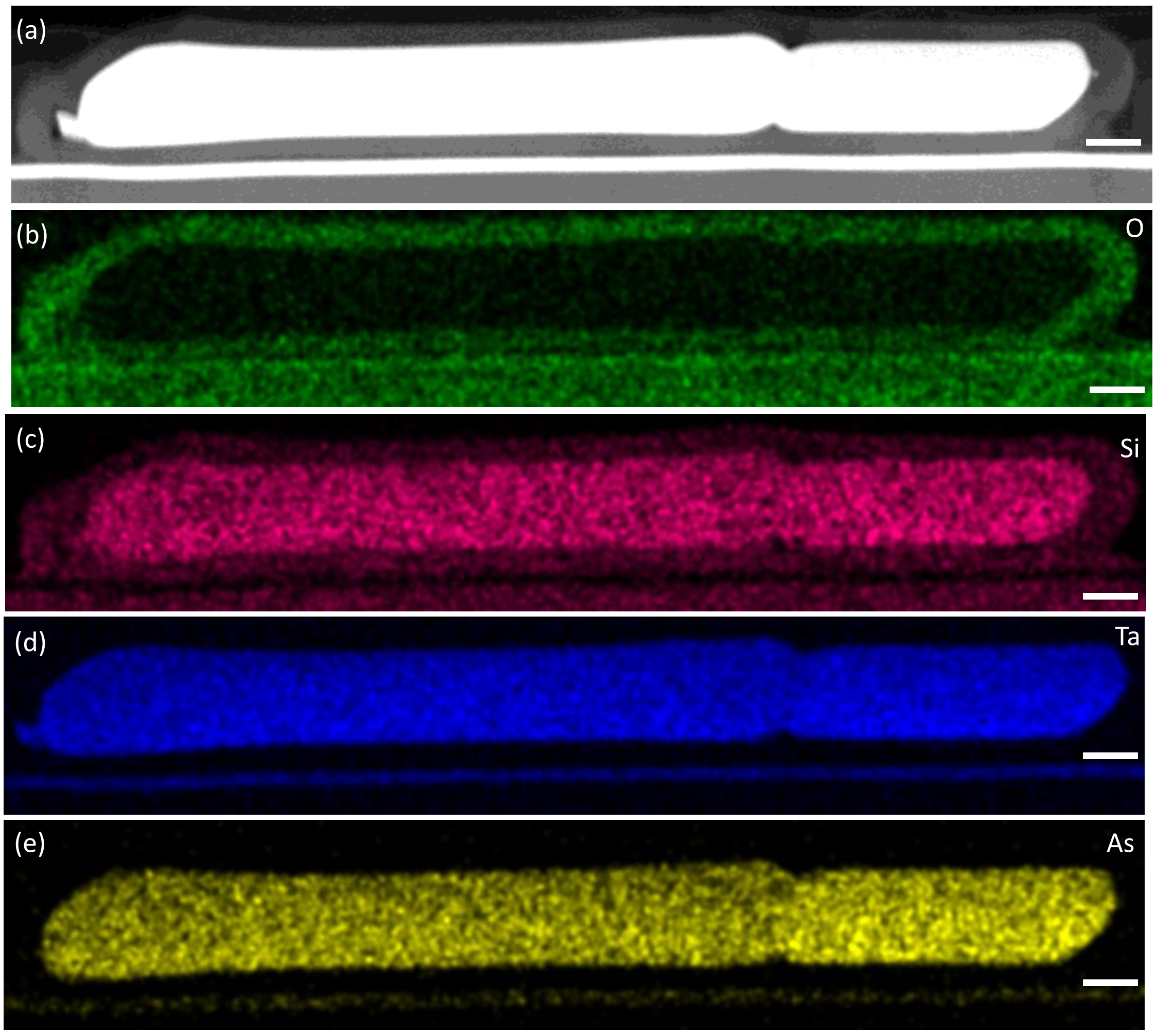

**Figure S7.** (a) HAADF-STEM image of a NBs cross-section, showing a core-shell structure with a TaSiAs core consisting of two grains. (b-e) STEM-EDS elemental mapping showing oxygen (green), silicon (magenta), tantalum (blue), and arsenic (yellow) signals. Scale bar: 50 nm.

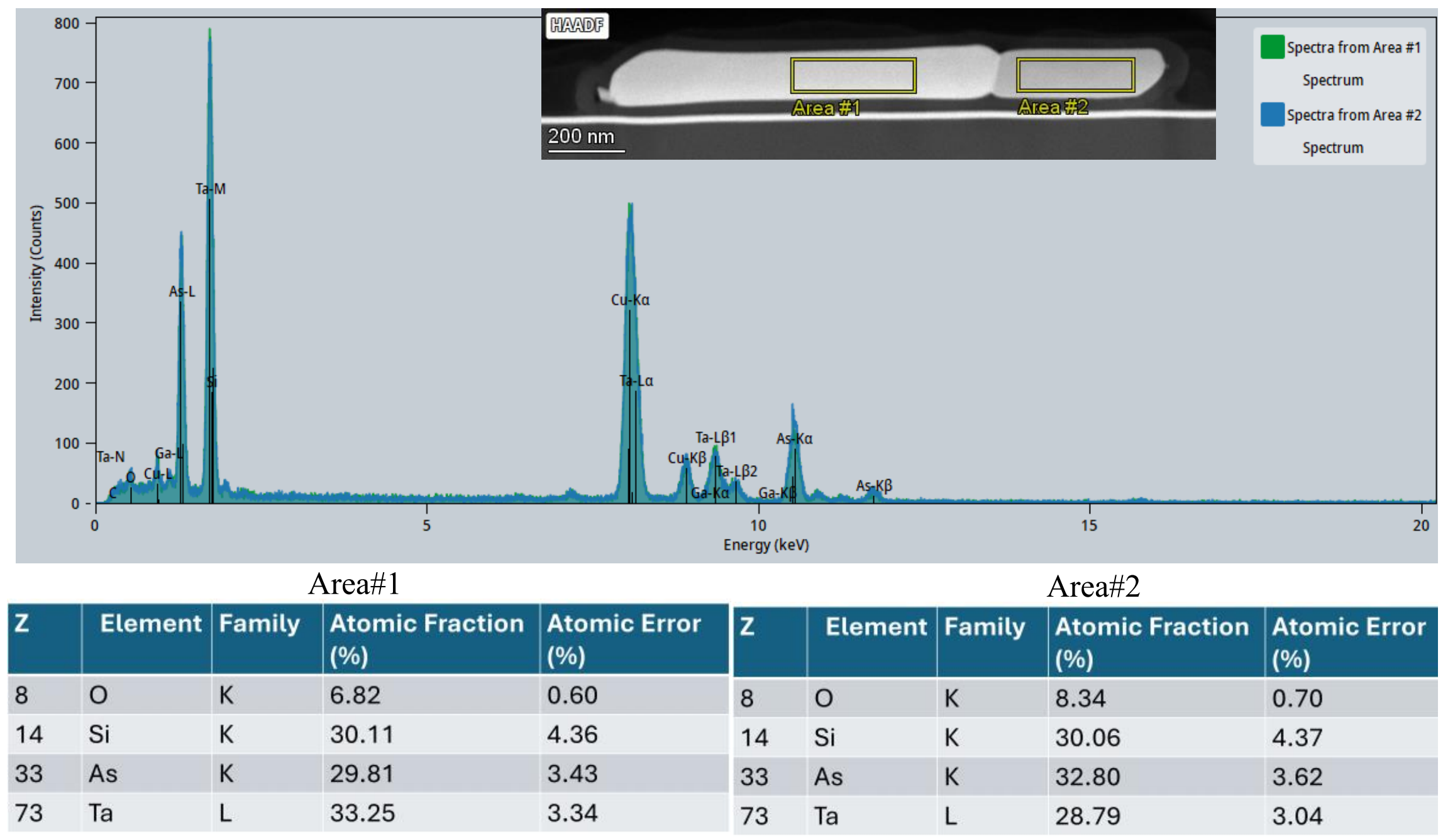

Area#1

| Z | Element | Family | Atomic Fraction (%) | Atomic Error (%) |
|---|---|---|---|---|
| 8 | O | K | 6.82 | 0.60 |
| 14 | Si | K | 30.11 | 4.36 |
| 33 | As | K | 29.81 | 3.43 |
| 73 | Ta | L | 33.25 | 3.34 |

Area#2

| Z | Element | Family | Atomic Fraction (%) | Atomic Error (%) |
|---|---|---|---|---|
| 8 | O | K | 8.34 | 0.70 |
| 14 | Si | K | 30.06 | 4.37 |
| 33 | As | K | 32.80 | 3.62 |
| 73 | Ta | L | 28.79 | 3.04 |

**Figure S8.** STEM-EDS spectra obtained from two different grains of the same NB TaSiAs composition.

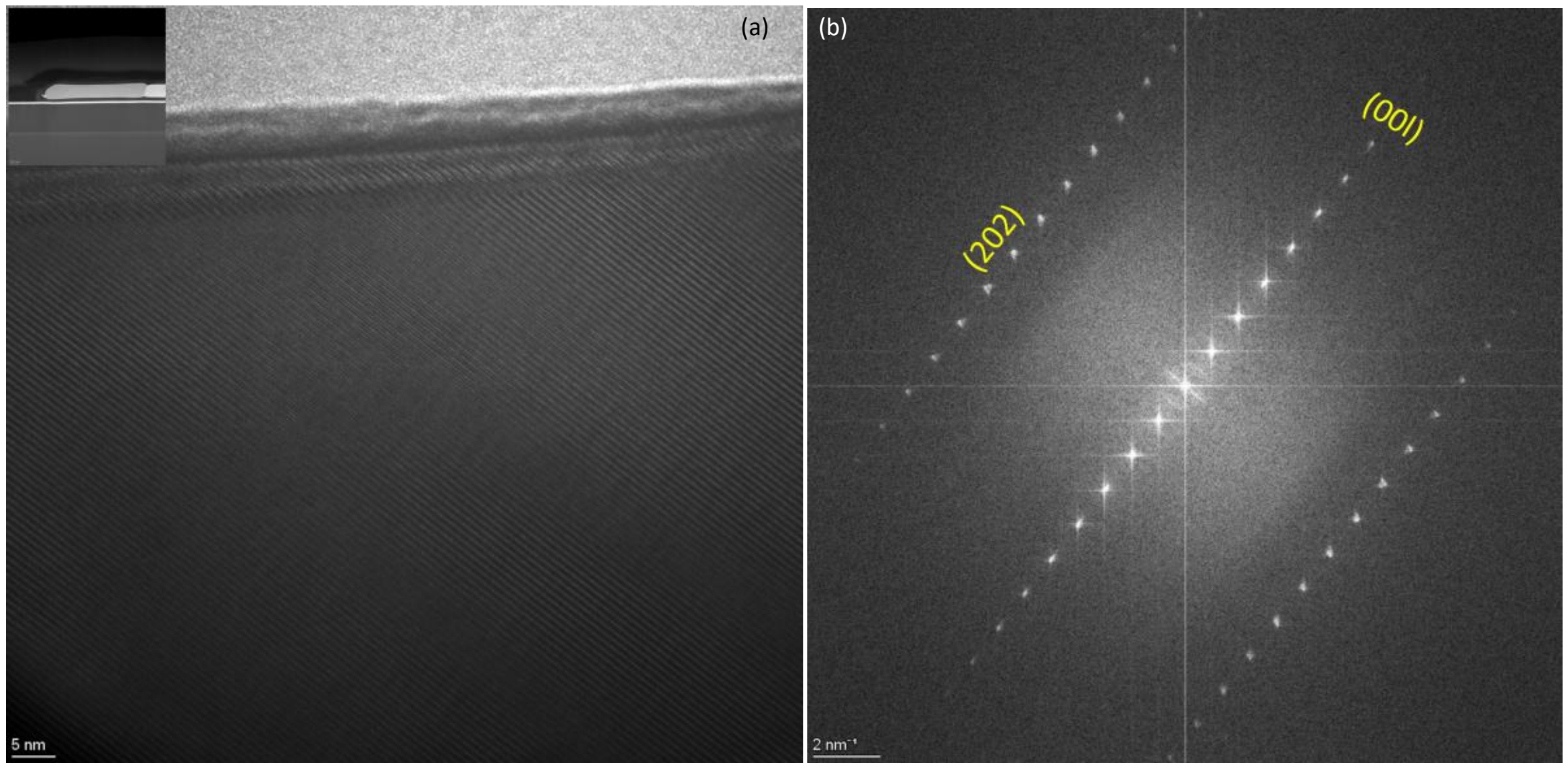

**Figure S9.** (a) HRTEM image obtained from the bigger grain of NB (inset), showing diagonally orientated layers of TaSiAs. (b) Corresponding FFT pattern confirming the <010> growth direction of the grain shown in the inset of **a**.

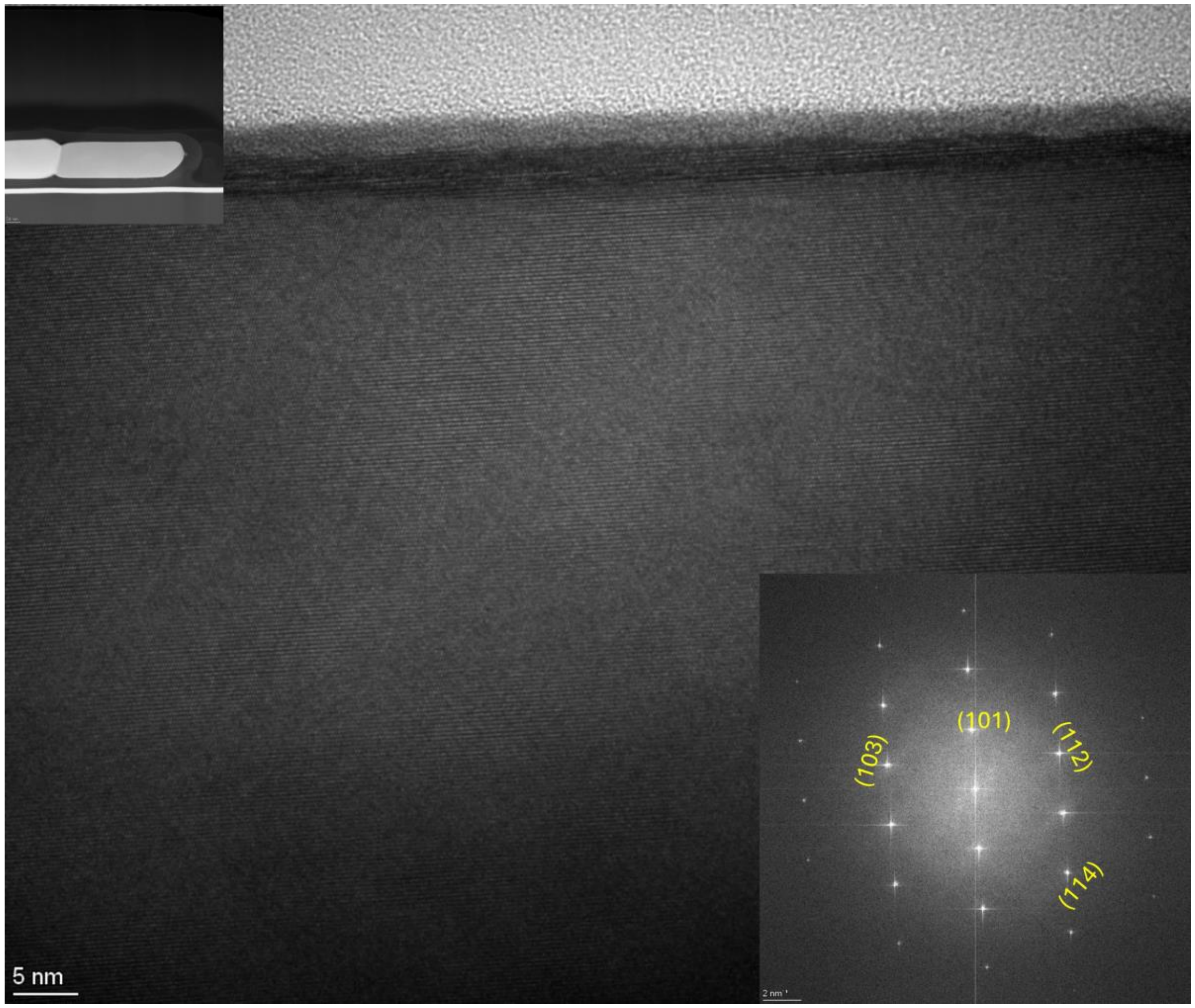

**Figure S10.** HRTEM image obtained from the smaller grain (inset) of the NB showing stacking of TaSiAs layers. Lower inset shows corresponding FFT pattern confirming a different <$\bar{1}\bar{1}$1> direction of the smaller grain.

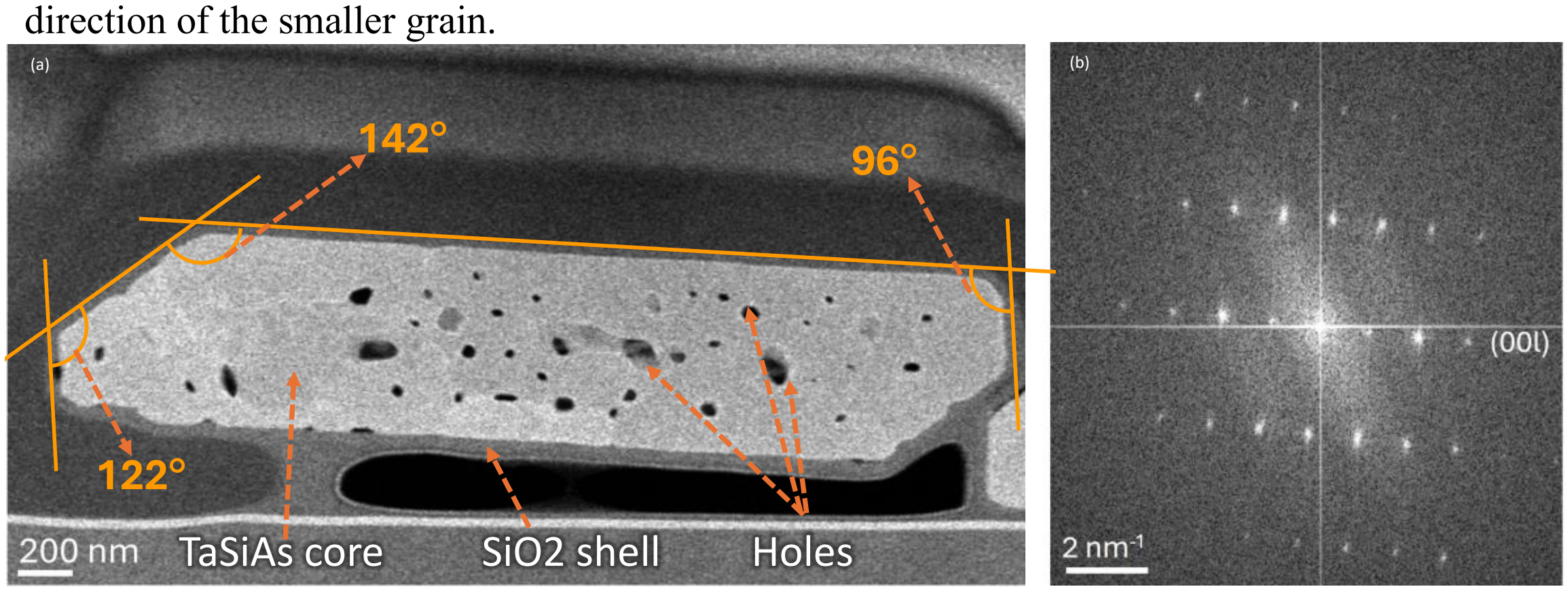

**Figure S11.** (a) HAADAF-STEM image of a thicker NB showing mosaic patterns indicative of different density regions of TaSiAs. The pseudofacet planes and the angles between them are shown in orange and are very close to those observed in $TaAs_2$ NWs (i.e. 141°, 121° and 97°). (b) Corresponding FFT pattern confirming a TaSiAs phase with an <010> growth direction.

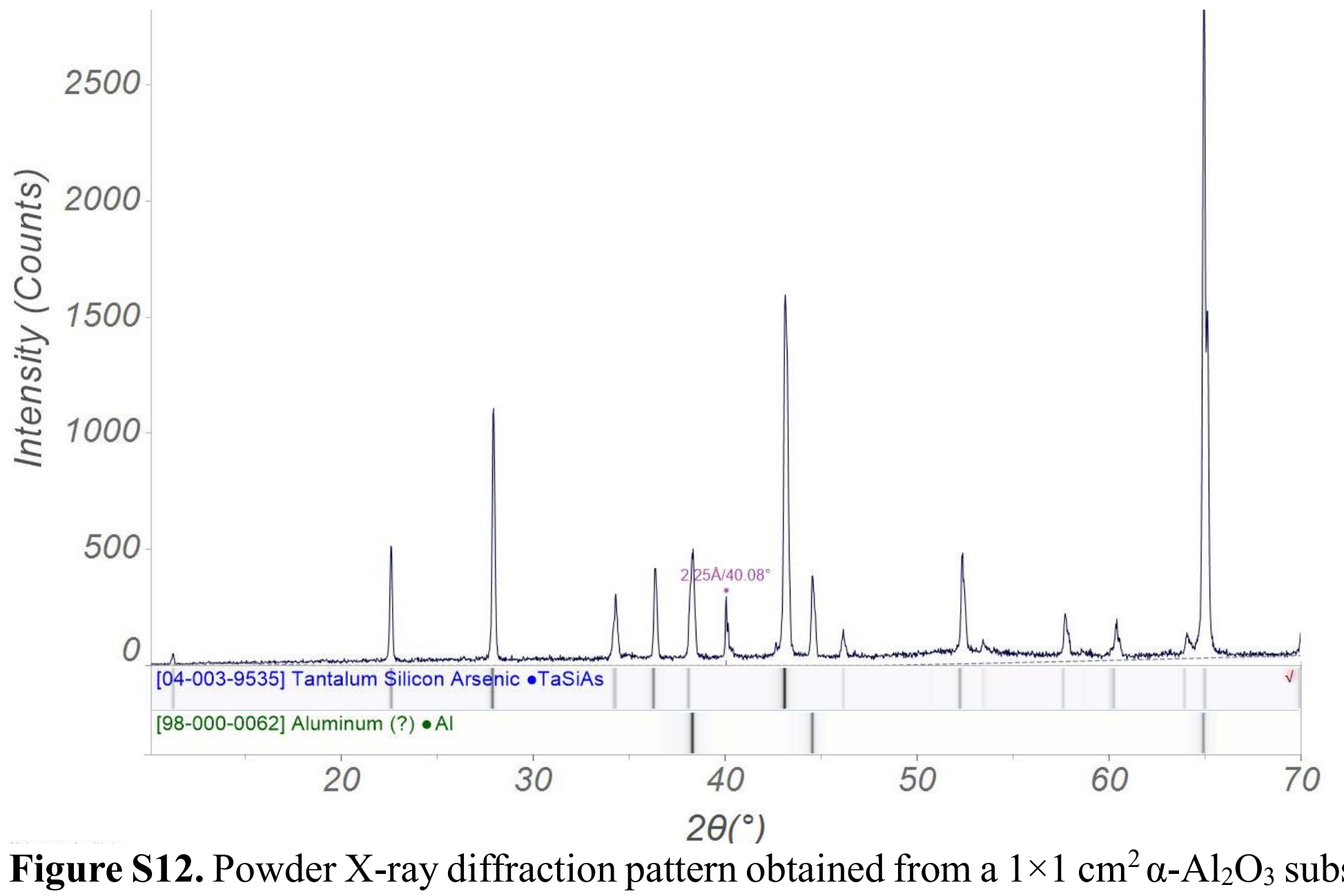

**Figure S12.** Powder X-ray diffraction pattern obtained from a 1×1 $cm^2$ α-$Al_2O_3$ substrate covered with NWs and NBs.

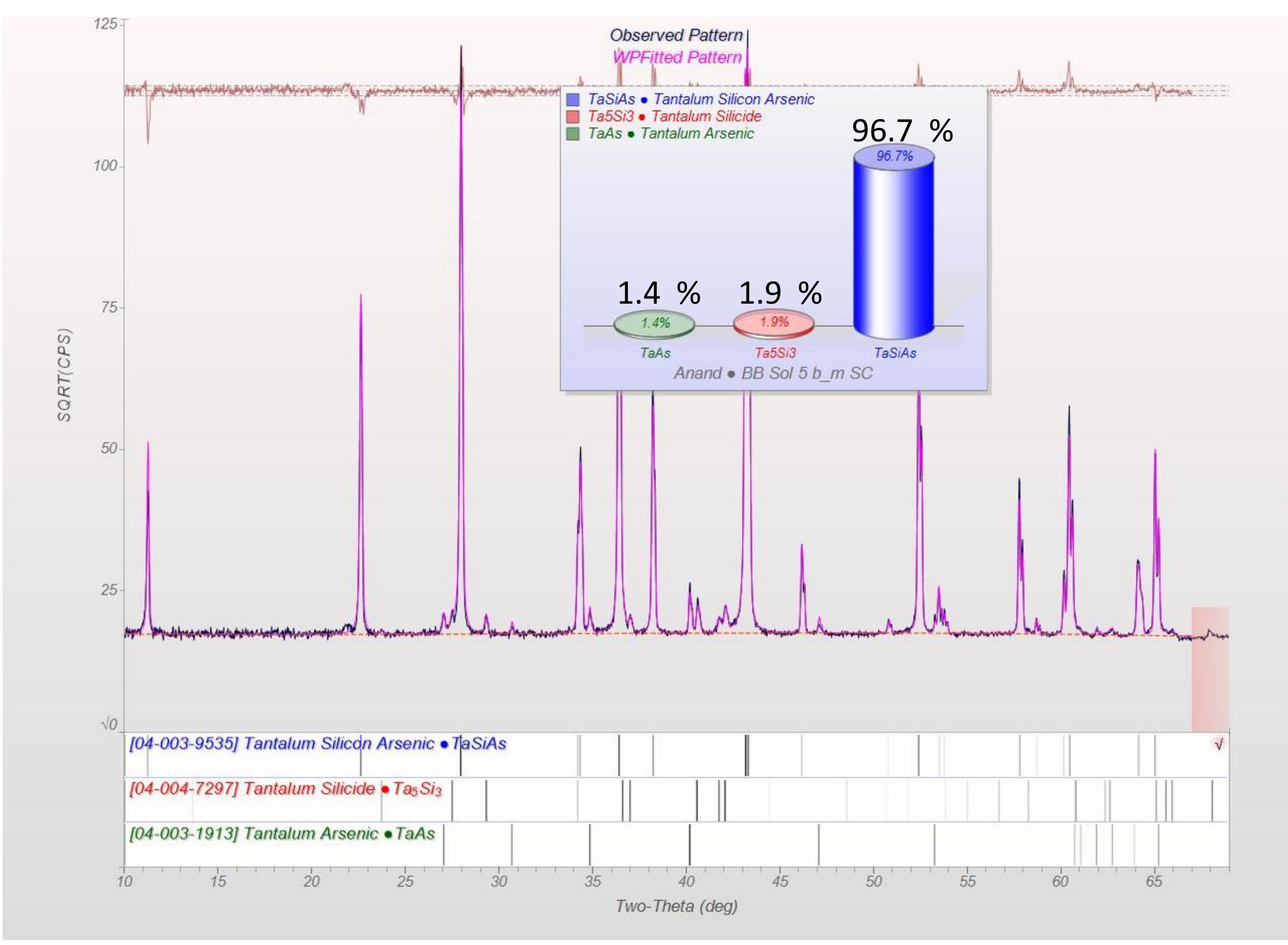

**Figure S13.** Powder X-ray diffraction pattern of polycrystalline TaSiAs powder. Diffraction pattern and relative contribution of other phases are shown in top and bottom inset.

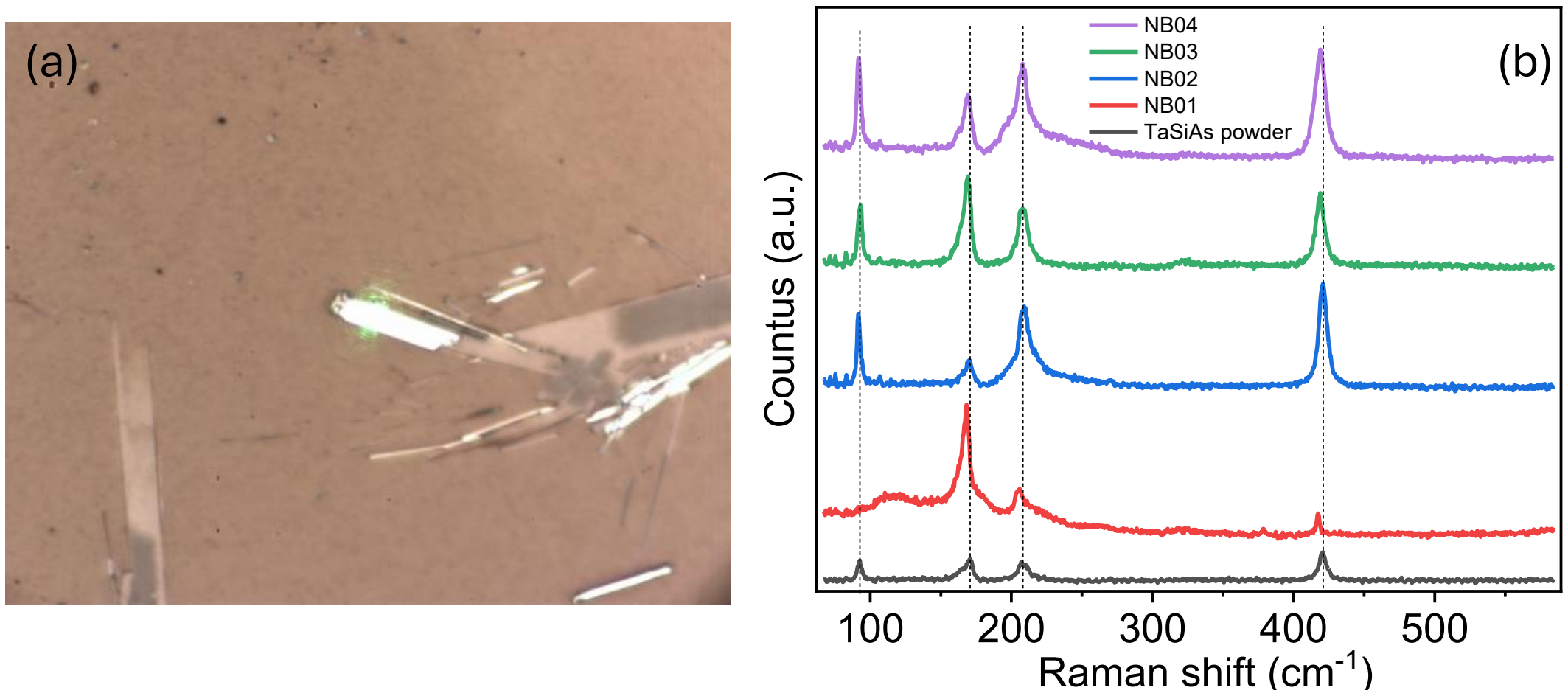

**Figure S14.** (a) An optical image showing dispersed NWs and NBs on a $Si/SiO_2$ substrate. Note that the yellow dotted circle highlights a focused spot of green laser (532 nm) on a NB sample. (b) Raman spectra of NBs, compared to that of reference TaSiAs powder.

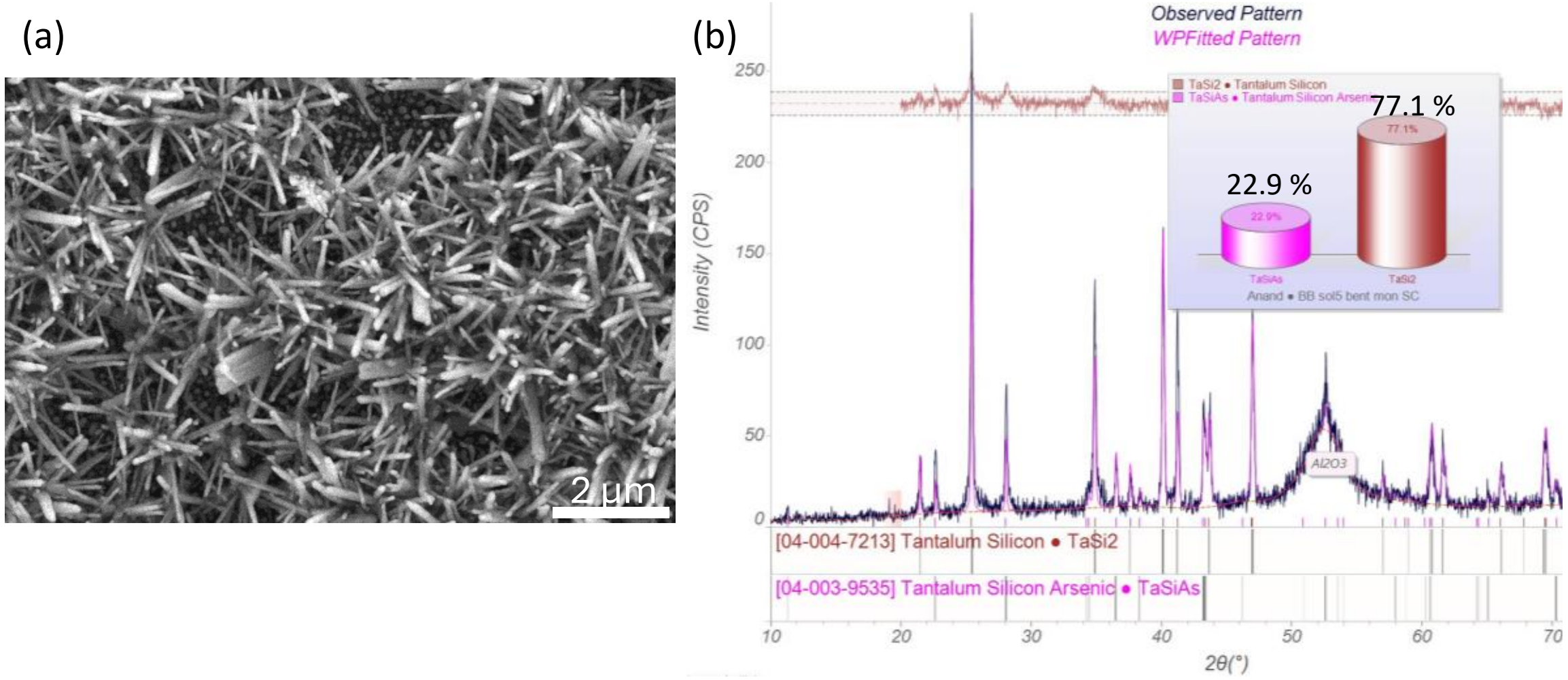

**Figure S15.** (a) SEM image of the reaction product obtained using Si powder instead of a Si substrate (with native oxide). (b) X-ray diffraction pattern of same sample, showing that the product is composed of $TaSi_2$ (≈77 %) and TaSiAs (≈23 %).

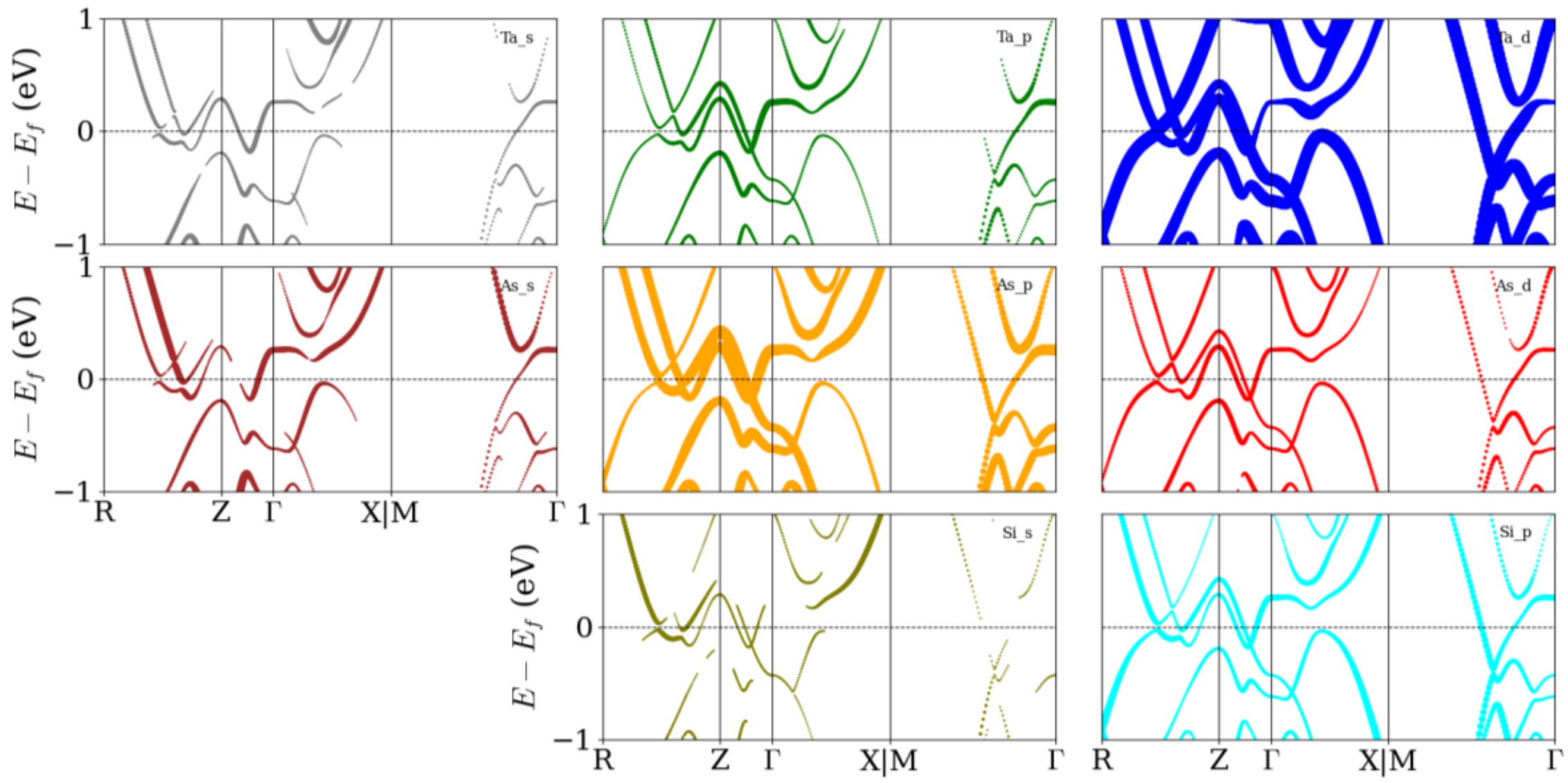

**Figure S16.** Orbital projected band structure of TaSiAs.

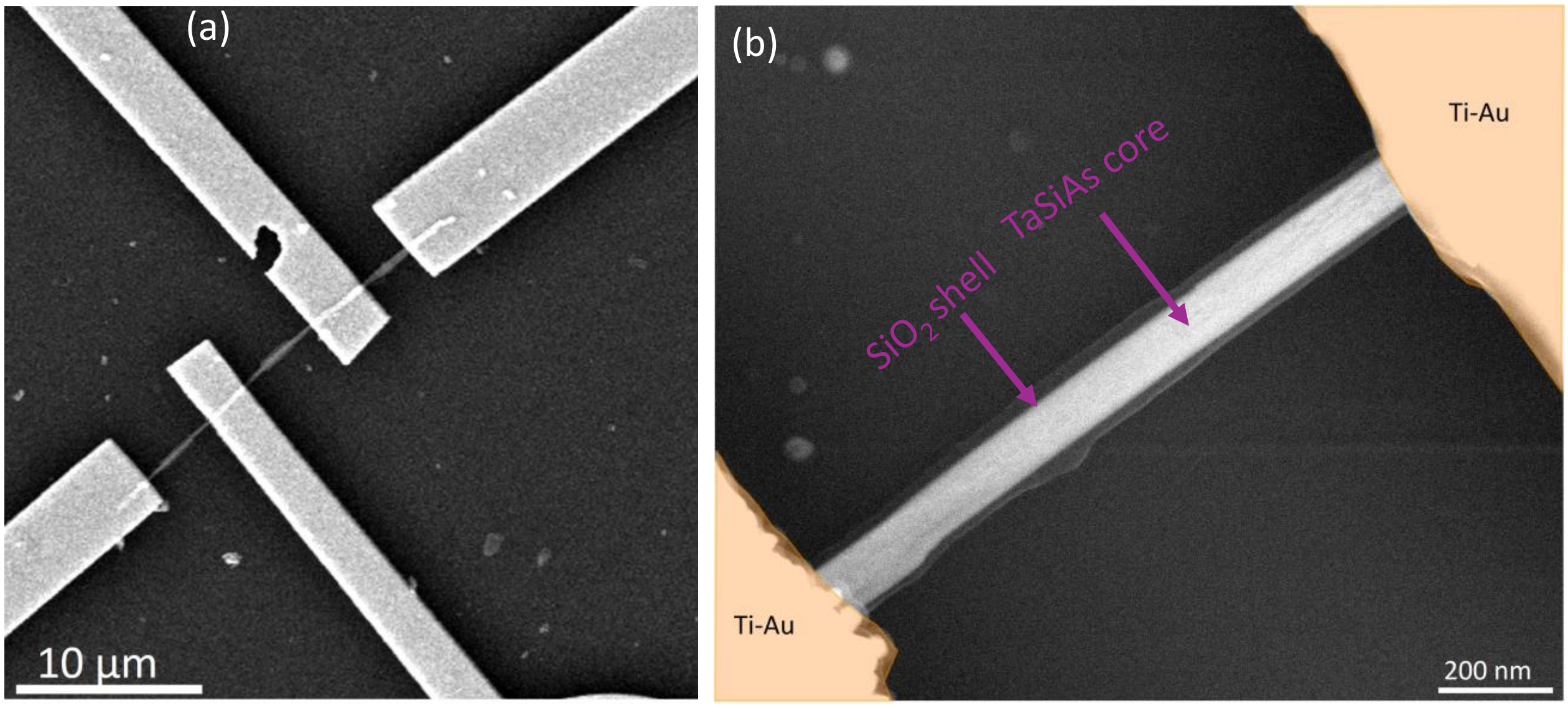

**Figure S17.** (a) SEM image of a four-terminal NW-device fabricated by local etching of a $SiO_2$ shell to make Ohmic contacts with the TaSiAs core. (b) High-magnification SEM image obtained from the portion of the NW in between the electrodes, displaying a $SiO_2$ shell encapsulated TaSiAs core. Note: Ti/Au electrodes in **b** are shown in false yellow color to highlight the electrodes.

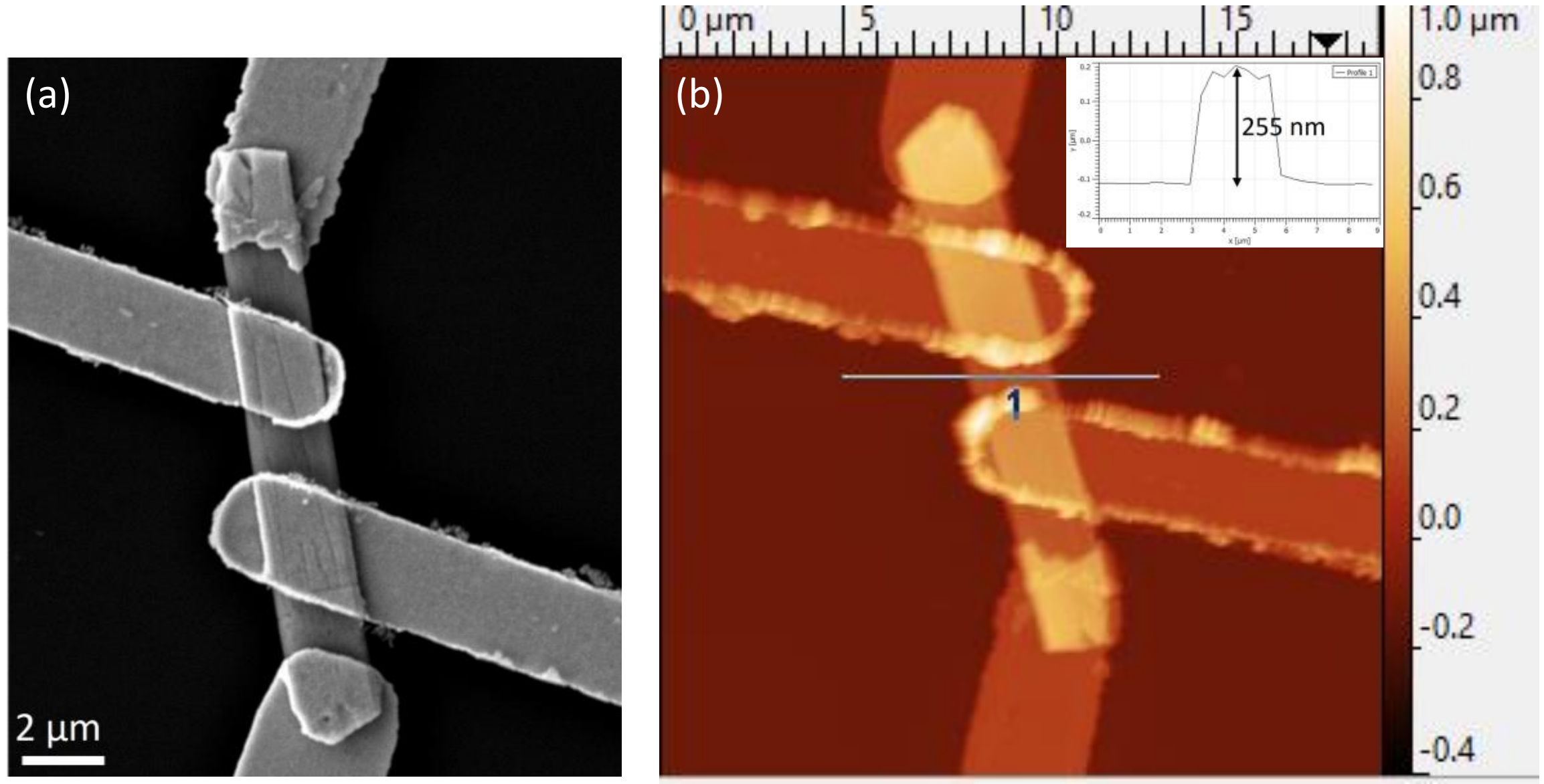

**Figure S18.** (a) SEM image of a NB device. (b) AFM image and height profile (inset) of the same device.

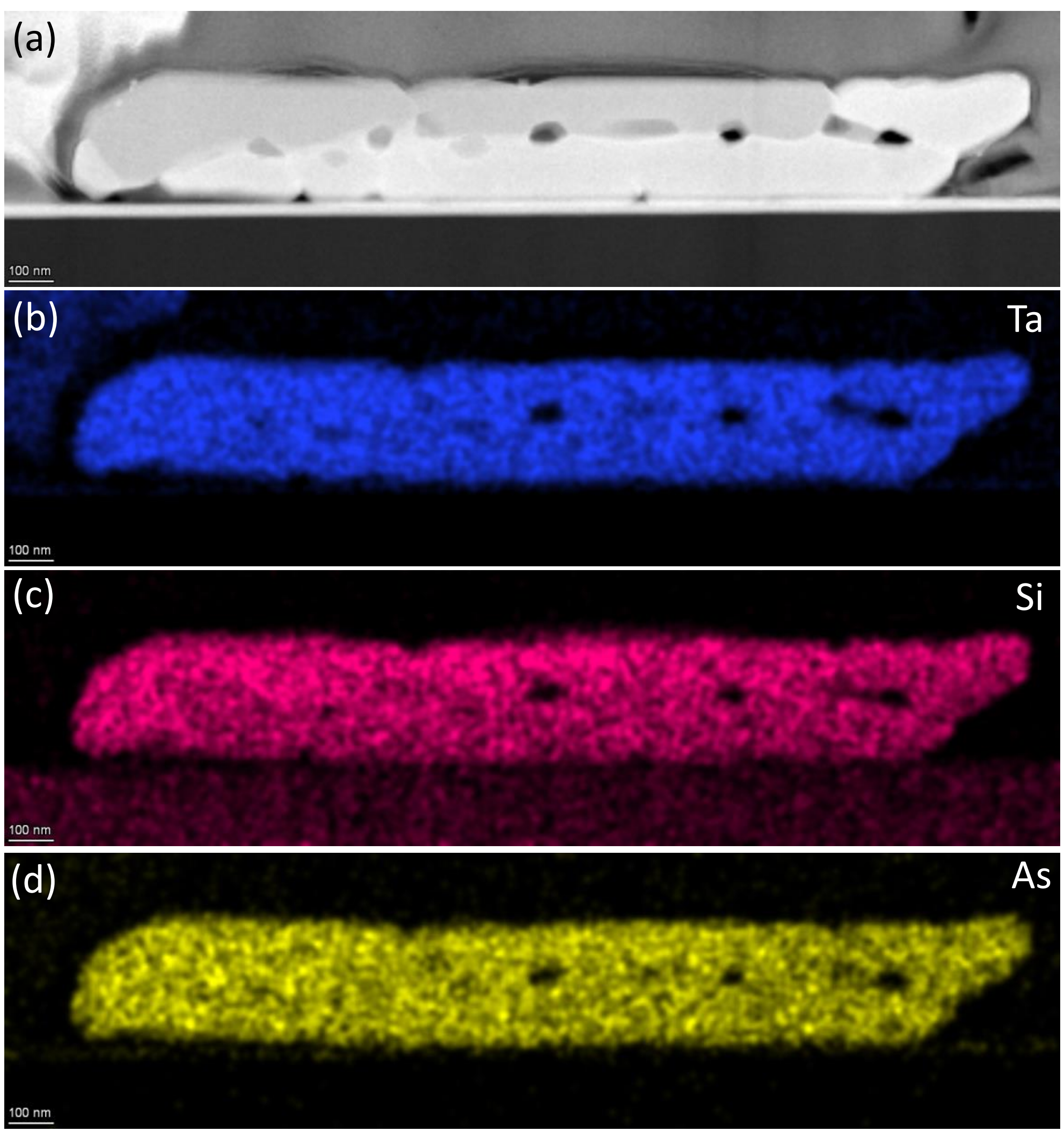

**Figure S19.** (a) HAADF-STEM image obtained from the cross-section of an NB01 device. (b-d) Corresponding STEM-EDS elemental mapping showing signals of Ta, Si, and As.

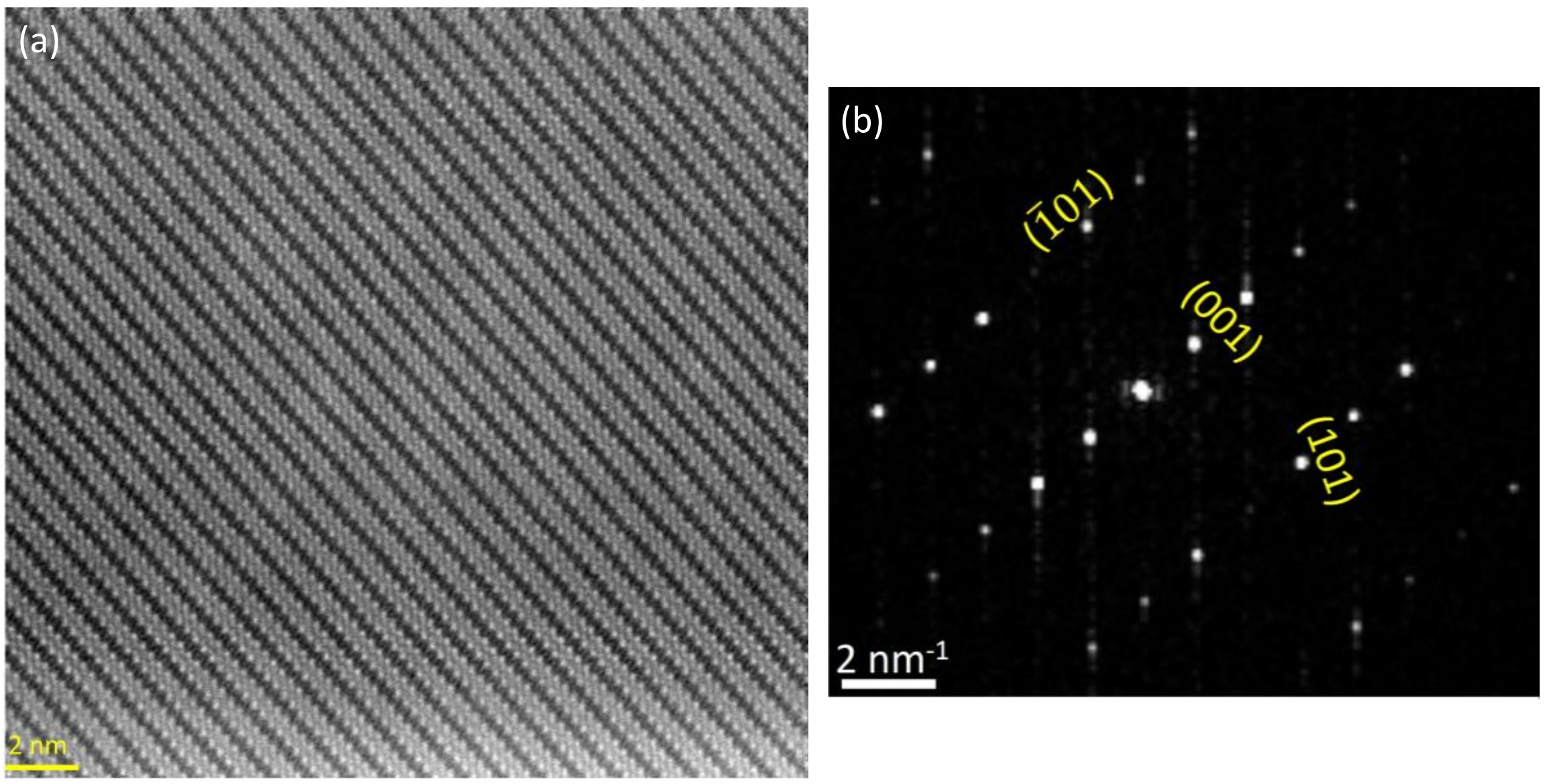

**Figure S20.** (a) High-resolution TEM image showing TaSiAs layers and their orientation in NB01 device. (b) Corresponding FFT pattern with identified planes.

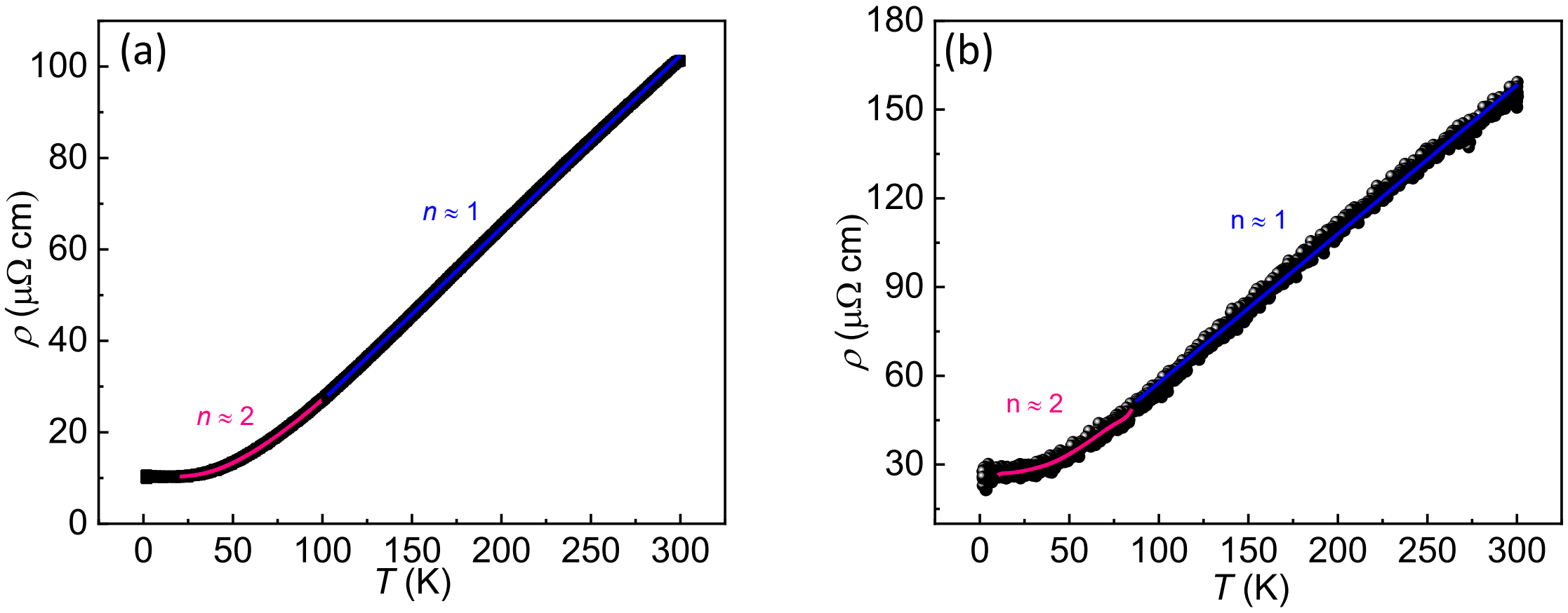

**Figure S21.** Resistivity as a function of temperature in a $d$ = 125 nm NW (a) and 255 nm thick NB01 (b). The red and blue lines represent fitting of data with $\rho\ (T) =\ \rho_0 + A\, T^n$.

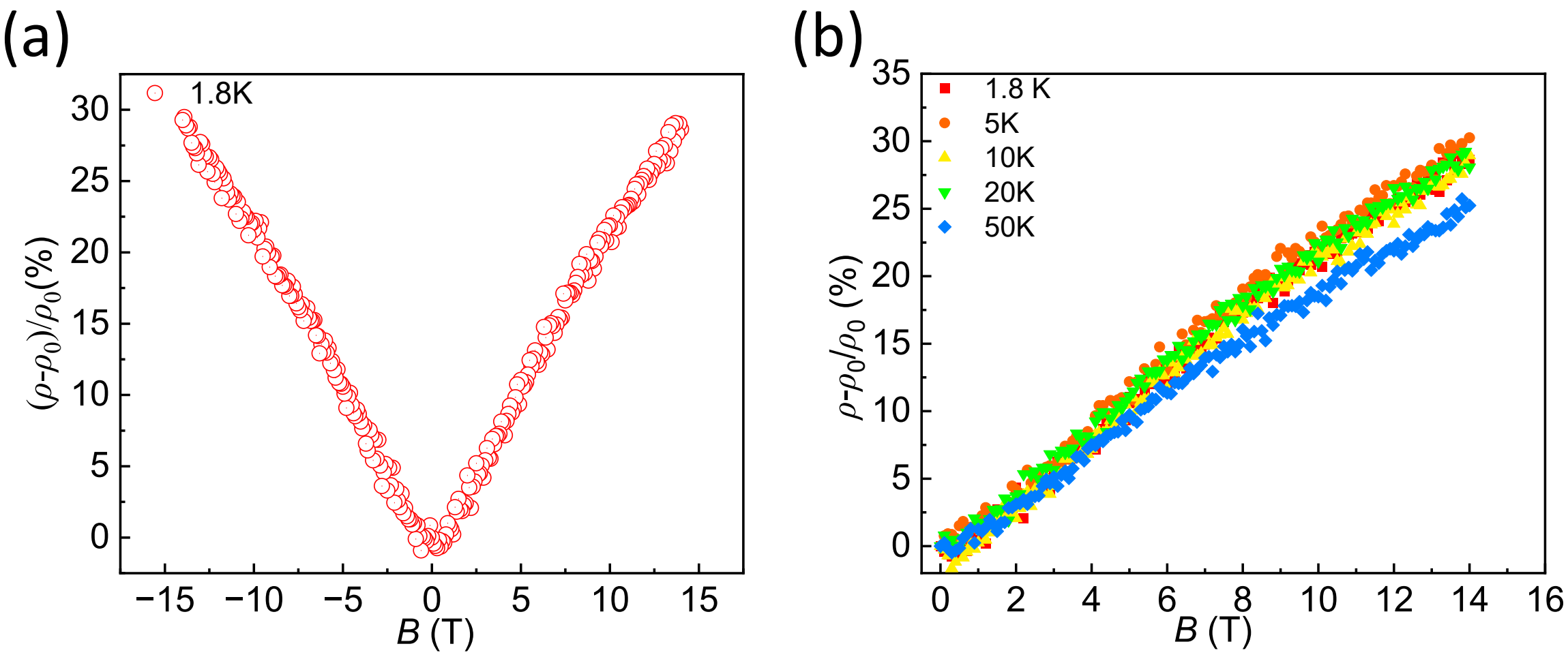

**Figure S22.** (a) Magnetoresistance, ($\rho$-$\rho_0$)/$\rho_0$, as a function of magnetic field strength ($B$) for a $d$ ≈ 90 nm NW in a two terminal device configuration. (b) Temperature dependence of MR for the same device.

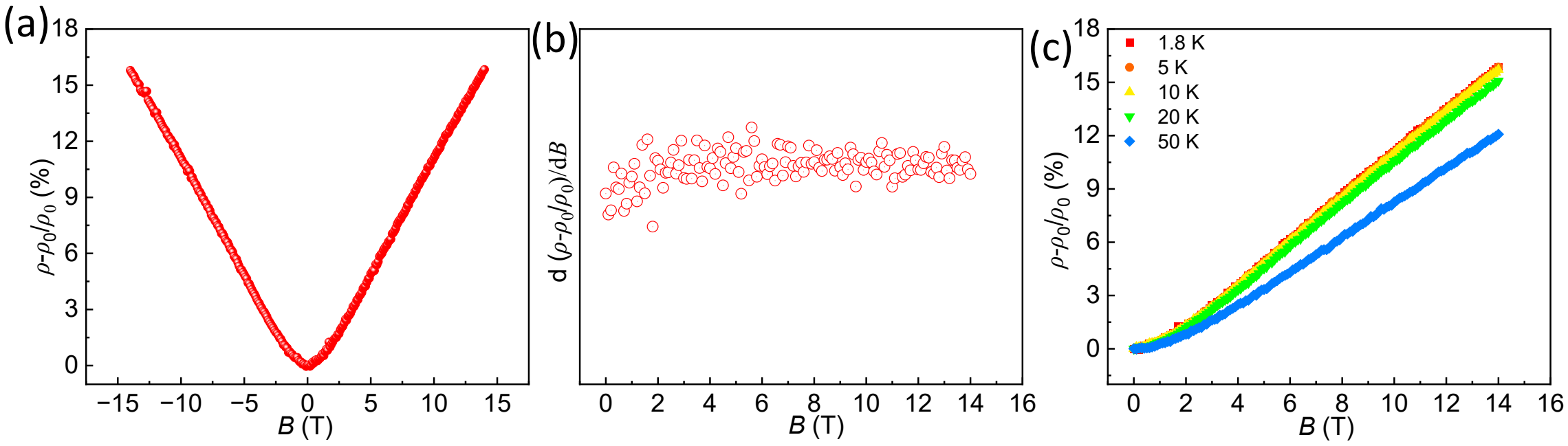

**Figure S23.** Magnetotransport features of a $d \sim 76$ nm NW device. (a) Magnetoresistance (MR), ($\rho$-$\rho_0$)/$\rho_0$, as a function of magnetic field ($B$) at 1.8 K. (b) First derivative of MR, d($\rho$-$\rho_0$)/$\rho_0$/d$B$, with respect to the field strength (B).(c) MR as a function of magnetic field ($B$) at different temperatures for the same $d \approx 76$ NW device.

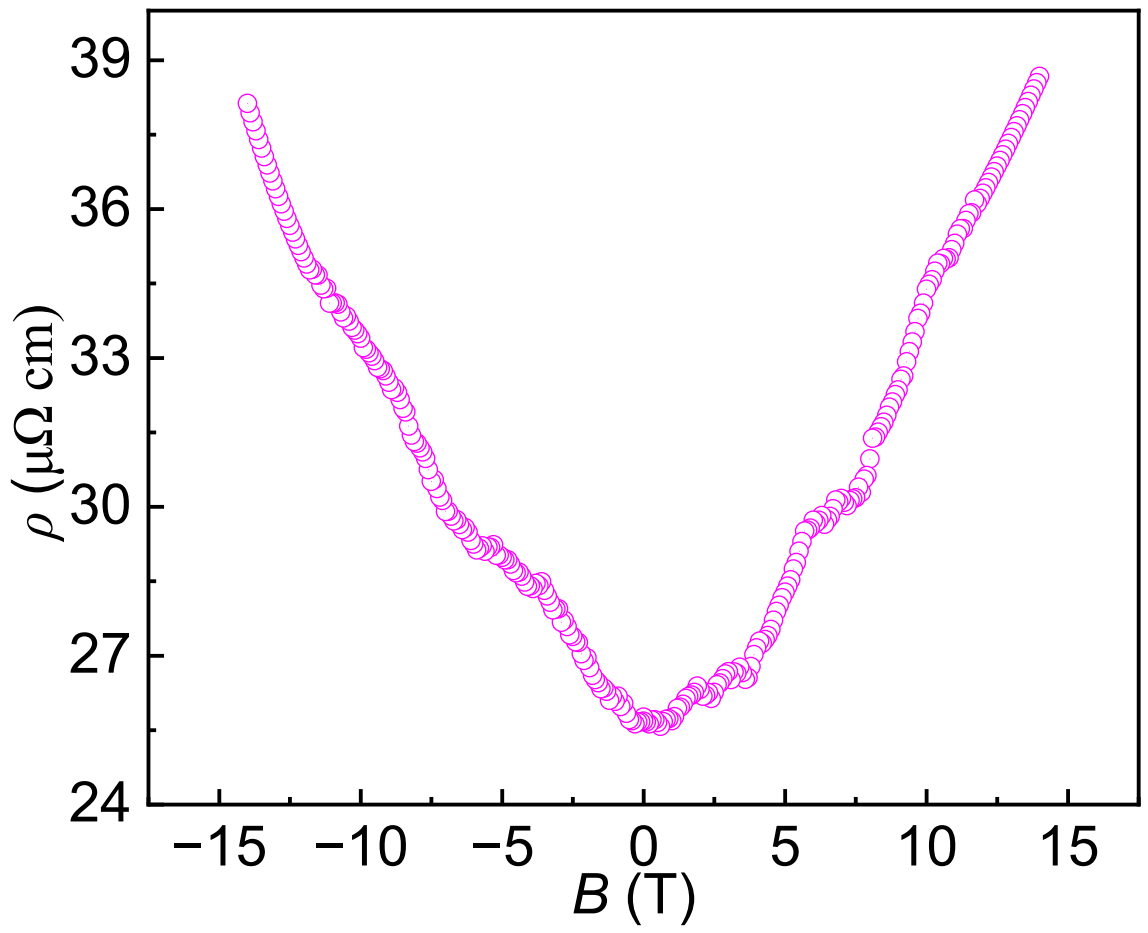

**Figure S24.** Resistivity as a function of magnetic field strength (*B*) for a NB device of thickness 255 nm.